\documentclass[preprint,floats,tightenlines,11pt,eqsecnum,showpacs,aps]{revtex4}
\usepackage{amsmath}
\usepackage{graphicx}
\begin{document}
\def\appls{\hbox{$<$\kern-.75em\lower 1.00ex\hbox{$\sim$}}}

\title{CONSISTENCY TESTS OF $\rho^0(770)-f_0(980)$ MIXING IN $\pi^- p \to \pi^-\pi^+n$}

\author{Miloslav Svec\footnote{electronic address: svec@hep.physics.mcgill.ca}}
\affiliation{Physics Department, Dawson College, Montreal, Quebec, Canada H3Z 1A4}
\date{June 10, 2015}

\begin{abstract}

Analytical solutions of the $S$- and $P$-wave subsystem in $\pi^- p \to \pi^- \pi^+ n$ and $\pi^+ n \to \pi^+ \pi^- p$ measured on polarized targets at CERN reveal evidence for $\rho^0(770)-f_0(980)$ spin mixing. We study the response of these analytical solutions to the presence of small $D$ wave amplitudes with helicity $\lambda \leq 1$ (Response analysis A) and $\lambda \leq 2$ (Response analysis B) which contaminate the input data. In both Response analyses the $\rho^0(770)-f_0(980)$ spin mixing effect is clearly consistent with the presence of the $D$-wave amplitudes provided they are not too large below 750 MeV. We also show that the $\rho^0(770)-f_0(980)$ mixing is consistent with isospin relations for the $S$-wave intensities measured in $\pi^- p \to \pi^- \pi^+ n$, $\pi^- p \to \pi^0 \pi^0 n$ and $\pi^+ p \to \pi^+ \pi^+ n$ processes.  We present a survey of moduli of the $S$-wave amplitudes and $S$-wave intensities from all amplitude analyses of the five measurements of $\pi^-p \to \pi^-\pi^+ n$ and $\pi^+ n \to \pi^+ \pi^- p$  on polarized targets. All analyses are in a remarkable agreement that shows a clear evidence for a resonant structure at $\rho^0(770)$ mass in the $S$-wave moduli and intensities in a broad confirmation of the $\rho^0(770)-f_0(980)$ spin mixing. We comment on our recent elastic and joint $\pi\pi$ phase-shift analyses of the CERN $\pi^-\pi^+$ and E852 $\pi^0\pi^0$ data and their agreement with the 1997 and 2002 Cracow Solutions, respectively. Our key observation is that all these solutions for the phase shift $\delta^0_S$ are consistent with the evidence for $\rho^0(770)-f_0(980)$ mixing documented by our survey. Together all these consistency results strengthen the experimental evidence for the $\rho^0(770)-f_0(980)$ spin mixing in $\pi^- p \to \pi^-\pi^+ n$ and are in agreement with recent theoretical expectations.

\end{abstract}

\maketitle

\tableofcontents

\newpage
\section{Introduction.}

The evidence for a rho-like state in the $S$-wave amplitudes in $\pi^- p \to \pi^- \pi^+n$ dates back to 1960's~\cite{hagopian63,islam64,patil64,durand65,baton65,gasiorowicz66} and was confirmed later in CERN measurements on polarized targets in $\pi^- p \to \pi^- \pi^+n$ at 17.2 GeV/c~\cite{donohue79,becker79a,becker79b,chabaud83,rybicki85,kaminski97,svec92c,svec96,svec97a} and in $\pi^+ n \to \pi^+ \pi^- p$ at 5.98 and 11.85~\cite{lesquen85,svec92a,svec92c,svec96,svec97a}. Additional evidence came from the ITEP data on $\pi^- p \to \pi^- \pi^+ n$ on polarized target at 1.78 GeV/c~\cite{alekseev98,alekseev99}. These findings were controversial because the measurements of $\pi^- p \to \pi^0 \pi^0 n$ at CERN in 1972~\cite{apel72} and at BNL in 2001~\cite{gunter01} found no evidence for the rho-like meson in the $S$-wave amplitudes. Using three different methods we show in a recent work~\cite{svec12b} that the rho-like resonance in the $S$-wave transversity amplitudes arises entirely from the contribution of the $\rho^0(770)$ resonance. In addition, there is a dip at the $f_0(980)$ mass in the $P$-wave aplitude $|L_d|^2$. These results present evidence for a $\rho^0(770)-f_0(980)$ mixing in $\pi^- p \to \pi^- \pi^+n$. Since there is no $P$-wave in $\pi^- p \to \pi^0 \pi^0 n$ this explains why there is no rho-like resonance observed in this process. The theoretical interpretation of the evidence for $\rho^0(770)-f_0(980)$ spin mixing is developed in Ref.~\cite{svec13a,svec13b} and applied to a new amplitude analysis of the CERN data using the spin mixing mechanism in Ref.~\cite{svec14a}.

In this work we strengthen the evidence for the $\rho^0(770)-f_0(980)$ mixing 
in two ways. First, we show that the observed mixing is not generated by the admixture of small $D$-wave amplitudes in the input data. The spin mixing effect is consistent with the presence of the $D$-wave amplitudes at low as well as at high momentum transfers. Second, we show that the $\rho^0(770)-f_0(980)$ mixing is consistent with the isospin relations between the observed  amplitudes in $\pi^- p \to \pi^- \pi^+ n$ and $\pi^- p \to \pi^0 \pi^0 n$. 

The paper is organized as follows. In Section II. we define the observables $t^L_M,p^L_M,r^L_M,q^L_M$ measured in $\pi^- p \to \pi^- \pi^+ n$ on polarized target when the polarization of the recoil nucleon is not observed. In Section III. we define new observables with definite nucleon transversity $\tau$ called $a_{k,\tau},k=1,15$ in terms of $t^L_M,p^L_M$ and express them in terms of partial wave transversity amplitudes $S_\tau$, $P_\tau$ and $D_\tau$. In Section IV. we discuss the solvability of this system of equations for the $S$, $P$ and $D$ wave amplitudes and focus on the response of the analytical solutions of the $S$- and $P$-wave subsystem to the presence of $D$-wave amplitudes. In both Response analyses the $\rho^0(770)-f_0(980)$ spin mixing effect is clearly consistent with the presence of the $D$-wave amplitudes provided they are not too large below 750 MeV. In Section V. we derive isospin relations between $S$-wave intensities in $\pi^- p \to \pi^- \pi^+ n$, $\pi^- p \to \pi^0 \pi^0 n$ and $\pi^+ p \to \pi^+ \pi^+ n$ processes and present experimental evidence for the consistency of the $\rho^0(770)-f_0(980)$ mixing with these relations. 

In Section VI. we survey the results for the moduli of the $S$-wave transversity amplitudes and for $S$-wave intensities from all amplitude analyses of five measurements of $\pi N \to \pi^-\pi^+ N$ on polarized targets and compare the latest results for relative phases. All analyses are in a remarkable agreement that shows a clear evidence for a resonant structure at $\rho^0(770)$ mass in the $S$-wave moduli and intensities in a broad confirmation of $\rho^0(770)-f_0(980)$ spin mixing. 

In Section VI. we also comment on our recent joint $\pi\pi$ phase-shift analysis of $\pi^-\pi^+$ and $\pi^0\pi^0$ production data~\cite{svec15a}. Our Solutions (1,1) joint and (2,2) joint for the phase-shift $\delta^0_S$ are in excellent agreement with the Solutions "up-flat" and "down-flat" from the Cracow analysis~\cite{kaminski02}, respectively. Our key observation is that all these phase-shifts $\delta^0_S$ are consistent with the evidence for the $\rho^0(770)-f_0(980)$ spin mixing in the $S$-wave transversity amplitudes from which they are ultimately derived. The paper closes with a Summary in Section VII.     

\section{The observables in $\pi^- p \to \pi^- \pi^+ n$ on polarized target.}

Consider the pion production process $\pi^- p \to \pi^- \pi^+ n$ with four-momenta $p_a+p_b = p_1+p_2+p_d$. The invariant mass of the dipion system is $m^2=(p_1+p_2)^2$. The angular distribution of the produced dipion system is described by the direction of $\pi^-$ in the two-pion center-of-mass system and its solid angle $\Omega = \theta, \phi$. When the polarization of the recoil nucleon is not measured the angular intensity takes the form~\cite{svec12b,lutz78,sakrejda84}
\begin{equation}
I(\Omega,\psi)=I_U(\Omega)+P_T\cos \psi I_C(\Omega)+P_T\sin \psi I_S(\Omega)+P_L I_L(\Omega)
\end{equation}
Here $\vec{P}=(P_x,P_y,P_z)=(P_T\sin \psi,P_T\cos \psi, P_L)$ is the target polarization vector where $P_T$ and $P_L$ are transverse and logitudinal polarization components perpendicular and parallel to the $z$-axis, respectively. The angle $\psi$ is the angle between $\vec{P_T}$ and the $y$-axis. In the laboratory system of the reaction the $+z$ axis has the direction opposite to the incident pion beam. The $+y$ axis is perpendicular to the scattering plane and has the direction of $\vec {p}_a \times \vec {p}_c$ where $p_c=p_1+p_2$. Dipion helicities are measured in the $t$-channel helicity frame  (dipion z-axis in direction of incident beam). The helicities of the nucleons are measured in the $s$-channel helicity frame~\cite{becker79a,lutz78}.

We shall use the parametrization of the angular components $I_U,I_C,I_S,I_L$ due to Lutz and Rybicki~\cite{lutz78,sakrejda84,becker79a,chabaud83,rybicki85}
\begin{eqnarray}
I_U(\Omega) & = & \sum \limits_{L,M} t^L_M ReY^L_M(\Omega)\\
\nonumber
I_C(\Omega) & = & \sum \limits_{L,M} p^L_M ReY^L_M(\Omega)\\
\nonumber
I_S(\Omega) & = & \sum \limits_{L,M} r^L_M ImY^L_M(\Omega)\\
\nonumber
I_L(\Omega) & = & \sum \limits_{L,M} q^L_M ImY^L_M(\Omega)
\end{eqnarray}
The parametrization (2.2) assumes $P$-parity conservation. In terms of density matrix elements the parameters $t,p,r.q$ read~\cite{lutz78,sakrejda84}
\begin{eqnarray}
t^L_M & = & \sum \limits_J \sum \limits_{J'\lambda'} K^{LM}_{JJ'\lambda'}Re \bigl(R^0_u \bigr)^{J,J'}_{M+\lambda',\lambda'}\\
\nonumber
p^L_M & = & \sum \limits_J \sum \limits_{J'\lambda'} K^{LM}_{JJ'\lambda'}Re \bigl(R^0_y \bigr)^{J,J'}_{M+\lambda',\lambda'}\\
\nonumber
r^L_M & = & \sum \limits_J \sum \limits_{J'\lambda'} K^{LM}_{JJ'\lambda'}Im \bigl(R^0_x \bigr)^{J,J'}_{M+\lambda',\lambda'}\\
\nonumber
q^L_M & = & \sum \limits_J \sum \limits_{J'\lambda'} K^{LM}_{JJ'\lambda'}Im \bigl(R^0_z \bigr)^{J,J'}_{M+\lambda',\lambda'}
\end{eqnarray}
where
\begin{equation}
K^{LM}_{JJ'\lambda'}=(-1)^{\lambda'} \sqrt{\frac{(2J+1)(2J'+1)}{4\pi (2L+1)}}<JJ'00|L0><JJ'M+\lambda'-\lambda'|LM>
\end{equation}
General expressions for the full set of density matrix elements $(R^j_k)^{JJ'}_{\lambda \lambda'},k=u,y,x,z$ including recoil nucleon polarization $j=1,2,3$ in terms of the unnatural and natural exchange transversity amplitudes $U^J_{\lambda,\tau}$ and $N^J_{\lambda,\tau}$ are given in Ref.~\cite{svec13a,lutz78}. Here $\tau=+\frac{1}{2},-\frac{1}{2}=up(u),down(d)$ is the target nucleon transversity. It is these amplitudes which are most suitable for the amplitude analysis of meaurements on both polarized and unpolarized targets.

\section{The $S$-, $P$- and $D$-wave subsystem.}

The $S$-,$P$- and $D$-wave subsystem is described by parameters $t,p,r,q$ for $L \leq 4$ and $M \leq 4$. The CERN measurements on transversely polarized target did not measure the parameters $q^L_M$. Expressions for $t,p,r$ in terms of the transversity amplitudes for $L\leq 4$ and $M \leq 2$ corresponding to $J\leq 2$ and $\lambda \leq 1$ were given by Lutz and Rybicki in Ref.~\cite{lutz78}. Expressions for $t,p,r$ for $L\leq 4$ and $M \leq 4$ corresponding to $J\leq 2$ and $\lambda \leq 2$ were given by Sakrejda in Ref.~\cite{sakrejda84}. 

In this work we focus on the parameters $t^L_M$ and $p^L_M$. These parameters organize themselves into two groups: $t^L_M+p^L_M$ are expressed in terms of bilinear terms $Re(A_uB^*_u)$ with transversity $up$, while  $t^L_M-p^L_M$ are expressed in terms of bilinear terms $Re(A_dB^*_d)$ with transversity $down$. We define the following convenient set of observables $a_{i,\tau}. i=1,15$
\begin{eqnarray}
a_{1,\tau}=\sqrt{\pi}(t^0_0 \pm p^0_0) & , & a_{2,\tau}=\sqrt{\pi}(t^2_0 \pm p^2_0)\sqrt{5}\\
\nonumber
a_{3,\tau}=\sqrt{\pi}(t^2_2 \pm p^2_2)\bigl(-\sqrt{\frac{5}{6}} \bigr) & , & a_{4,\tau}=\sqrt{\pi}(t^1_0 \pm p^1_0)\frac{1}{2}\\
\nonumber
a_{5,\tau}=\sqrt{\pi}(t^2_1 \pm p^2_1)\bigl(\frac{1}{2}\sqrt{\frac{5}{6}} \bigr) & , & a_{6,\tau}=\sqrt{\pi}(t^1_1 \pm p^1_1)\bigl 
(\frac{1}{2}\sqrt{\frac{1}{2}}\bigr )
\end{eqnarray}
\begin{eqnarray}
a_{7,\tau}=\sqrt{\pi}(t^3_0 \pm p^3_0)\bigl(\frac{1}{6}\sqrt{\frac{35}{3}} \bigr) & , & a_{8,\tau}=\sqrt{\pi}(t^3_1 \pm p^3_1)\bigl(\frac{1}{8}\sqrt{\frac{35}{3}}\bigr)\\
\nonumber
a_{9,\tau}=\sqrt{\pi}(t^3_2 \pm p^3_2)\bigl(\frac{1}{2}\sqrt{\frac{7}{6}} \bigr) & , & a_{10,\tau}=\sqrt{\pi}(t^4_0 \pm p^4_0)\frac{7}{2}\\
\nonumber
a_{11,\tau}=\sqrt{\pi}(t^4_1 \pm p^4_1)\bigl(\frac{7}{4}\sqrt{\frac{1}{35}} \bigr) & , & a_{12,\tau}=\sqrt{\pi}(t^4_2 \pm p^4_2)\bigl(\frac{7}{2}\sqrt{\frac{1}{10}}\bigr)
\end{eqnarray}
\begin{eqnarray}
a_{13,\tau}=\sqrt{\pi}(t^3_3 \pm p^3_3)\bigl(\frac{\sqrt{7}}{3} \bigr) & , & a_{14,\tau}=\sqrt{\pi}(t^4_3 \pm p^4_3)\bigl( \sqrt{\frac{7}{5}} \bigr)\\
\nonumber
a_{15,\tau}=\sqrt{\pi}(t^4_4 \pm p^4_4)\bigl( \sqrt{\frac{14}{5}} \bigr)
\end{eqnarray}
In (3.1)-(3.3) $\tau=u$ for the $+$ sign and $\tau=d$ for the $-$ sign. Next we express the observables $a_{i,\tau}$ in terms of $S$-, $P$- and $D$-wave amplitudes defined as follows
\begin{eqnarray}
\begin{array} {lll}
U^0_{0,\tau}=S_\tau &  &  \\
U^1_{0,\tau}=L_\tau & U^1_{1,\tau}=U_\tau & N^1_{1,\tau}=N_\tau\\
U^2_{0,\tau}=D^0_\tau & U^2_{1,\tau}=D^U_\tau & U^2_{2,\tau}=D^{2U}_\tau\\
N^2_{1,\tau}=D^N_\tau & N^2_{2,\tau}=D^{2N}_\tau &   \\
\end{array}
\end{eqnarray}
For the purposes of our analysis we shall split the observables $a_{i,\tau}$ into three parts
\begin{equation}
a_{i,\tau}=c_{i,\tau}+d_{i,\tau}+e_{i,\tau}
\end{equation}
where $c_{i,\tau}$ involve only $S$- and $P$-wave amplitudes, $d_{i,\tau}$ involve terms with $D$-wave amplitudes with only helicity $\lambda \leq 1$, and $e_{i,\tau}$ involve terms with $D$-wave amplitudes with $\lambda=2$ (rank 2 amplitudes). The expressions for the $D$-wave terms $d_{i,\tau}$ and $e_{i,\tau}$ in terms of the transversity amplitudes are given in the Table I. The expressions for $c_{i,\tau}$ read as follows
\begin{eqnarray}
c_{1,\tau} & = & |S_\tau|^2+|L_\tau|^2 +|U_\tau|^2+|N_\tau|^2 \\
\nonumber
c_{2,\tau} & = & 2|L_\tau|^2-|U_\tau|^2-|N_\tau|^2 \\
\nonumber
c_{3,\tau} & = & |N_\tau|^2-|U_\tau|^2 \\ 
\nonumber
c_{4,\tau} & = & |L_\tau||S_\tau|\cos \Phi(L_\tau S^*_\tau) \\
\nonumber
c_{5,\tau} & = & |L_\tau||U_\tau|\cos \Phi(L_\tau U^*_\tau) \\
\nonumber
c_{6,\tau} & = & |U_\tau||S_\tau|\cos \Phi(U_\tau S^*_\tau) 
\end{eqnarray}
where the cosines of relative phases
\begin{equation}
\cos \Phi(A_\tau B^*_\tau)=\cos(\Phi(A_\tau)-\Phi(B_\tau))
\end{equation}
All $c_{i,\tau}=0$ for $i=7,15$.

\begin{table}
\caption{$D$-wave contributions $d_{i,\tau}$ and $e_{i,\tau}$ to the observables $a_{i,\tau}$ corresponding to $D$-wave transversity amplitudes with helicities $\lambda \leq 1$ and $\lambda \leq 2$, respectively. The transversity index $\tau$ is omitted for the sake of brevity and the bilinear terms $AB^* \equiv Re(AB^*)$. Table from Ref.~\cite{lutz78,sakrejda84}.}
\begin{tabular}{|c|c|c|}
\toprule
$a_{i,\tau}$ & $d_{i,\tau}$ & $e_{i,\tau}$ \\
\colrule
$a_1$ & $|D^0|^2+|D^U|^2+|D^N|^2$ & $|D^{2U}|^2+|D^{2N}|^2$ \\
$a_2$ & $2\sqrt{5}D^0S^*+\frac{5}{7}(2|D^0|^2+|D^U|^2+|D^N|^2)$ & $-\frac{10}{7}(|D^{2U}|^2+|D^{2N}|^2)$ \\
$a_3$ & $\frac{5}{7}(|D^N|^2-|D^U|^2)$ & $-2\sqrt{\frac{5}{3}} SD^{2U*}+\frac{20}{7}\sqrt{\frac{1}{3}}D^0 D^{2U*}$ \\
$a_4$ & $\sqrt{\frac{4}{5}}D^0L^*+\sqrt{\frac{3}{5}}(D^UU^*+D^NN^*)$ & 0 \\
$a_5$ & $\sqrt{\frac{5}{3}}D^US^*+\frac{5}{7}\sqrt{\frac{1}{3}}D^UD^{0*}$ & 
$\frac{5}{7}(D^UD^{2U*}+D^ND^{2N*})$ \\
$a_6$ & $\sqrt{\frac{3}{5}}D^UL^*-\sqrt{\frac{1}{5}}D^0U^*$ & $\sqrt{\frac{3}{5}}(UD^{2U*}+ND^{2N*})$ \\
\colrule
$a_7$ & $D^0L^*-\sqrt{\frac{1}{3}}(D^UU^*+D^NN^*)$ & 0 \\
$a_8$ & $D^UL^*+\sqrt{\frac{3}{4}}D^0U^*$ & $-\frac{1}{4}(UD^{2U*}+ND^{2N*})$ \\
$a_9$ & $D^UU^*-D^NN^*$ & $LD^{2U*}$ \\
$a_{10}$ & $3|D^0|^2-2(|D^U|^2+|D^N|^2)$ & $\frac{1}{2}(|D^{2U}|^2+|D^{2N}|^2)$ \\
$a_{11}$ & $D^UD^{0*}$ & $-\frac{1}{2}\sqrt{\frac{1}{7}}(D^UD^{2U*}+D^ND^{2N*})$ \\
$a_{12}$ & $|D^U|^2-|D^N|^2$ & $\sqrt{3}D^0D^{2U*}$ \\
\colrule
$a_{13}$ & 0 & $UD^{2U*}-ND^{2N*}$ \\
$a_{14}$ & 0 & $D^UD^{2U*}-D^ND^{2N*}$ \\
$a_{15}$ & 0 & $|D^{2U}|^2-|D^{2N}|^2$ \\

\botrule
\end{tabular}
\label{Table I.}
\end{table}

\section{Consistency of $\rho^0(770)-f_0(980)$ mixing with the $D$-wave amplitudes.}

\subsection{Assessment of the data and solvability}

The full $S$-,$P$- and $D$-wave system $a_{i,\tau},i=1,15$ with $d_{i,\tau}\neq0$ and $e_{i,\tau}\neq0$ is not analytically solvable even when additional information on $r^L_M$ and $q^L_M$ is added. To obtain the moduli and relative phases of the amplitudes $\chi^2$ fits to the data are needed bin by bin~\cite{rybicki85}. Assuming $D^{2U}=D^{2N}=0$ all $e_{i,\tau}=0$. The truncated $S$-,$P$- and $D$-wave system $a_{i,\tau},i=1,12$ with $d_{i,\tau}\neq0$ is in general not analytically solvable. However assuming phase coherence of unnatural exchange amplitudes
\begin{eqnarray}
\Phi(U_\tau)-\Phi(L_\tau) & = & \pi\\
\nonumber
\Phi(D^U_\tau)-\Phi(D^0_\tau) & = & \pi
\end{eqnarray}
the simplified system is analytically solvable~\cite{lutz78,chabaud83}. Finally, assuming all $d_{i,\tau}=0,e_{i,\tau}=0$ the $S$- and $P$-wave subsystem is analytically solvable~\cite{lutz78,chabaud83,svec12b} for the moduli and relative phases of the $S$- and $P$-wave transversity amplitudes.

The evidence for $\rho^0(770)-f_0(980)$ mixing comes from the amplitude analyses of CERN data on polarized target at low $|t|$ below 1080 MeV assumng $S$- and $P$-wave dominance and $c_{i,\tau}=a_{i,\tau}$~\cite{svec12b}. Ideally we would like to separate the $S$- and $P$-wave amplitudes from the $D$-wave amplitudes to ensure that the $\rho^0(770)-f_0(980)$ mixing is not generated by the assumption $c_{i,\tau}=a_{i,\tau}$. To asses the CERN data on $\pi^- p \to \pi^- \pi^+ n$ from the point of view of evidence for  $\rho^0(770)-f_0(980)$ mixing the Table II. summarizes the measurements below 1080 MeV.

\begin{table}
\caption{The CERN measurements of $\pi^- p \to \pi^- \pi^+ n$ on unpolarized and polarized target at 17.2 GeV/c below dipion mass 1080 MeV.}
\begin{tabular}{|c|c|c|c|c|c|c|c|}
\toprule
Exp. & Observables & $m[MeV]$ & $|t|[(GeV/c)^2]$ & $L,M$ & $J=2,\lambda \leq 1$ & $J=2,\lambda=2$ & Ref. \\ 
\colrule
1 & $t^L_M$ & $300-1080$ & $0.00<|t|<0.15$ & $L\leq6,M\leq2$ & yes & no & \cite{grayer74} \\
2 & $t^L_M,p^L_M,r^L_M$ & $580-960$ & $0.005<|t|<0.20$ & $L\leq2,M\leq2$ & no & no & \cite{becker79b,chabaud83} \\
3 & $t^L_M,p^L_M,r^L_M$ & $960-1080$ & $0.005<|t|<0.20$ & $L\leq4,M\leq2$ & yes & no & \cite{becker79b,chabaud83} \\
4 & $t^L_M,p^L_M,r^L_M$ & $580-1080$ & $0.20<|t|<1.0$ & $L\leq4,M\leq4$ & yes & yes & \cite{rybicki85} \\

\botrule
\end{tabular}
\label{Table II.}
\end{table}

While the measurement 1 on unpolarized target presents evidence for small $D$-wave amplitudes with helicity $\lambda \leq 1$, the measurements 2 on polarized target are not sensitive to such amplitudes below 960 MeV. Measurements 1 and 2 at low $|t|$ below 960 MeV thus cannot be separately used to separate $S$-and $P$-wave amplitudes from the $D$-wave amplitudes. However the two measurements could be combined for observables with $L\leq4,M\leq2$ if we set $p^L_M=r^L_M=0$ for $L=3,4$ and $M\leq2$ in measurement 2, and if we assume the phase coherence (4.1) to solve analytically the approximate $S$-,$P$- and $D$-wave system.

Above 960 MeV the measurements 3 on polarized target at low $|t|$ are sensitive to the $D$-wave amplitudes with $\lambda \leq 1$ and the amplitude analyses~\cite{becker79b,chabaud83,kaminski02} separate $S$-, $P$- and $D$-wave amplitudes for $m \geq 980$ MeV. No measurement at low $|t|$ has detected the $\lambda =2$ $D$-wave amplitudes. However the measurement 4 on polarized target shows that these amplitudes are present at high $|t|$ even below 960 MeV. In this case the $S$- and $P$-wave amplitudes can be separated and present evidence for $\rho^0(770)-f_0(980)$ mixing~\cite{rybicki85} (Section IV.C).

\subsection{Response analysis at low $t$}

The $S$- and $P$-wave subsystem (3.6) is formally solvable for the moduli and cosines of relative phases in terms of the unknown observables $c_{i,\tau},i=1,6$. With
\begin{equation}
c_{i,\tau}=a_{i,\tau}-b_{i,\tau}=a_{i,\tau}-d_{i,\tau}-e_{i,\tau}
\end{equation}
and omitting the index $\tau$ for the sake of brevity, the solution reads
\begin{eqnarray}
|S|^2=c_1+c_2-3|L|^2 & = & a_1+a_2-3|L|^2 -b_1-b_2\\
\nonumber
|U|^2=|L|^2-\frac{1}{2} (c_2+c_3) & = & |L|^2-\frac{1}{2} (a_2+a_3)-\frac{1}{2} (b_2+b_3)\\
\nonumber
|N|^2=|L|^2-\frac{1}{2} (c_2-c_3) & = & |L|^2-\frac{1}{2} (a_2-a_3)-\frac{1}{2} (b_2-b_3)\\
\cos \Phi(LS^*)=\frac{c_4}{|L||S|} & = & \frac{a_4-b_4}{|L||S|}\\
\nonumber
\cos \Phi(LU^*)=\frac{c_5}{|L||u|} & = & \frac{a_5-b_5}{|L||u|}\\
\nonumber
\cos \Phi(US^*)=\frac{c_6}{|U||S|} & = & \frac{a_6-b_6}{|U||S|}
\end{eqnarray}
The phase condition $\Phi(LS^*)=\Phi(LU^*)+\Phi(US^*)$ implies a cosine condition
\begin{equation}
\cos \Phi(LS^*)^2 +\cos \Phi(LU^*)^2 +\cos \Phi(US^*)^2-2\cos \Phi(LS^*)\cos \Phi(LU^*)\cos \Phi(US^*)=1
\end{equation}
Substituting from (4.4) and (4.3) we obtain a qubic equation for $|L|^2$~\cite{lutz78,svec12b}. 

Assuming $c_{i,\tau}=a_{i,\tau}$ we neglect the unknown $D$-wave contributions $b_{i,\tau}$ but the system is analytically solvable with two physical solutions for the amplitudes with both transversities. Instead of solving the $S$-, $P$- and $D$-wave subsystem our strategy is to examine the response of the solutions of the equations (4.3),(4.4) and (4.5) to assumed $D$-wave contributions $b_{i,\tau}$. We want to find out if the solutions with $b_{i,\tau}=0,i=1,6$ change so much for small $D$-wave amplitudes that the $\rho^0(770)-f_0(980)$ mixing "disappears". If this "disappearance" does not happen, then we can be confident that the observed $\rho^0(770)-f_0(980)$ mixing is a genuine effect not generated by our assumption $c_{i,\tau}=a_{i,\tau}$. 

We perform two types of response analysis. In Response Analysis A we assume $D^{2U}_\tau=D^{2N}_\tau=0$ so that $b_{i,\tau}=d_{i,\tau}$. In Response Analysis B these amplitudes are no longer vanishing. In both cases the response analysis is feasible only when the system of equations (4.3),(4.4) and (4.5) remains analytically solvable. This solvability requires that the phases of $D$-wave amplitudes decouple from the $S$- and $P$-wave amplitudes such that
\begin{eqnarray}
Re(D^0_\tau S^*_\tau)=0\\
\nonumber
d_{4,\tau}=d_{5,\tau}=d_{6,\tau}=0\\
\nonumber
e_{3,\tau}=e_{5,\tau}=e_{6,\tau}=0
\end{eqnarray}
In both cases the critical amplitude to watch is the Solution 2 for the $S$-wave amplitude $|S_d|^2$.

\subsubsection{Response analysis A}

In this analysis we assume
\begin{eqnarray}
b_{1,\tau} & = & |D^0_\tau|^2+|D^U_\tau|^2+|D^N_\tau|^2\\
\nonumber
b_{2,\tau} & = & \frac{5}{7}\bigl(2|D^0_\tau|^2+|D^U_\tau|^2+|D^N_\tau|^2 \bigr)\\
\nonumber
b_{3,\tau} & = & \frac{5}{7}\bigl(|D^N_\tau|^2-|D^U_\tau|^2 \bigr)
\end{eqnarray}
For dipion masses $m>980$ MeV we know the $D$-wave intensities $I(A)=|A_u|^2+|A_d|^2$, $A=D^0,D^U,D^N$ from the amplitude analysis of the CERN measurement 3~\cite{chabaud83}. We linearly extrapolate these intensities from their values $I_2(A)$ at $m_2=990$ MeV to value $I_1(A)=T I_2(A)$ at $m_1=590$ MeV where the fraction $T$ defines the slope parameter. The extrapolated intensities at mass $m$ are 
\begin{equation}
I(A,m)= T I_2(A)+\frac{(1-T)I_2(A)} {m_2-m_1}(m-m_1)
\end{equation}
Below 980 MeV there is a fairly constant ratio of the moduli $|A_u|^2:|A_d|^2 \approx 1:3$ for all $S$-and $P$-wave amplitudes. Using this ratio we reconstruct the moduli of the $D$-wave amplitudes from the intensities
\begin{eqnarray}
|A_u(m)|^2=0.25I(A,m)F\\
\nonumber
|A_d(m)|^2=0.75I(A,m)F
\end{eqnarray}
where the factor $F$ accounts for the sudden decrease of the $D$-wave moments with $L=3,4$ below the $K\bar{K}$ threshold. We assume $F=0.50$. We vary the slope parameter $T$ in the range from 0.05 to 1.00 to estimate the $D$-wave amplitudes below 980 MeV. Above 980 MeV we used the amplitudes (4.9) calulated from the measured intensities of the analysis~\cite{chabaud83}.

The results for the critical Solution 2 of $|S_d|^2$ are shown in Figures 1 and 2 for $T=0.05, 0.10, 0.15, 0.20$ and $T=0.30,0.50, 0.75, 1.00$, respectively. We find that for $T < 0.70$ the $\rho^0(770)$ structure survives and is largely insensitive to the presence of the $D$-wave amplitudes for $T \lesssim 0.30$. However, below 749 MeV the analysis is compatible only with small $D$-wave amplitudes for $T \lesssim 0.30$, which is an expected result from the data of measurements 1.

\subsubsection{Response analysis B}

The measurements 3 on polarized target~\cite{chabaud83} indicate that $t^L_M=p^L_M=0$ for $L=3,4$ and $M=3,4$. From the Table I. we see that $t^4_4=p^4_4=0$ implies $|D^{2U}_\tau|^2 = |D^{2N}_\tau|^2$. Then
\begin{eqnarray}
b_{1,\tau} & = & |D^0_\tau|^2+|D^U_\tau|^2+|D^N_\tau|^2 +2|D^{2U}_\tau|^2\\
\nonumber
b_{2,\tau} & = & \frac{5}{7}\bigl(2|D^0_\tau|^2+|D^U_\tau|^2+|D^N_\tau|^2 
-4|D^{2U}_\tau|^2 \bigr)\\
\nonumber
b_{3,\tau} & = & \frac{5}{7}\bigl(|D^N_\tau|^2-|D^U_\tau|^2 \bigr)
\end{eqnarray}
We assume $|D^{2U}_\tau|^2$ is a fraction of $|D^U_\tau|^2$
\begin{equation}
|D^{2U}_\tau|^2 =T_2 |D^U_\tau|^2
\end{equation}
We examine the sensitivity of the solutions on $|D^{2U}_\tau|^2$ for small values of $T$. The results for the Solution 2 of $|S_d|^2$ are shown in Figures 3 and 4 for $T=0.05$ and $T=0.15$, respectively, for a broad range of $T_2$. Again, the $\rho^0(770)$ structure survives and is largely insensitive to $|D^{2U}_\tau|^2$. Above 1000 MeV the solutions require $|D^{2U}_\tau|^2=|D^{2N}_\tau|^2 \approx 0$ in excellent agreement with the absence of these amplitudes in the amplitude analyses~\cite{becker79b,chabaud83} of the measurements 3 up to dipion mass 1780 MeV.

\subsection{Amplitude analysis at high $t$}

Measurements 4 of $t^L_M,p^L_M,r^L_M$ with $L\leq 4, M\leq4$ at high momentum transfers $0.20 \leq |t| \leq 1.00$ (GeV/c$)^2$ and dipion mass $580 \leq m \leq 1500$ MeV revealed the presence of all $D$-wave amplitudes even below 960 MeV. Amplitude analysis~\cite{rybicki85} used $\chi^2$ fits of the data to determine the amplitudes. 

\begin{figure} [htp]
\includegraphics[width=12cm,height=10.5cm]{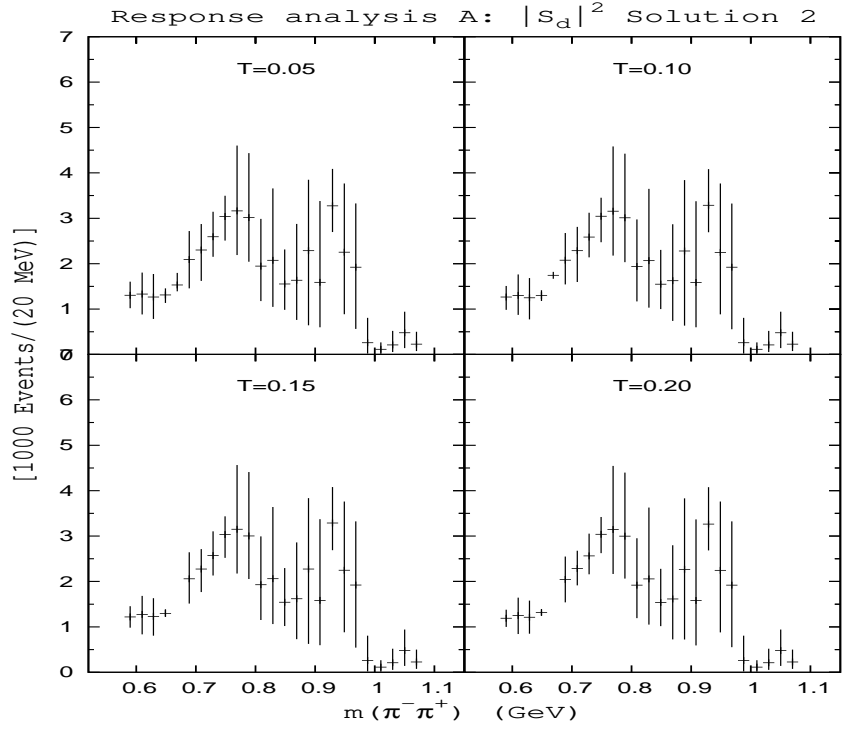}
\caption{Response of $S$-wave amplitude $|S_d|^2$ Solution 2 to
$|D^0_\tau|^2,|D^U_\tau|^2,|D^N_\tau|^2$ at low $T$.}
\label{Figure 1}
\end{figure}

\begin{figure} [hp]
\includegraphics[width=12cm,height=10.5cm]{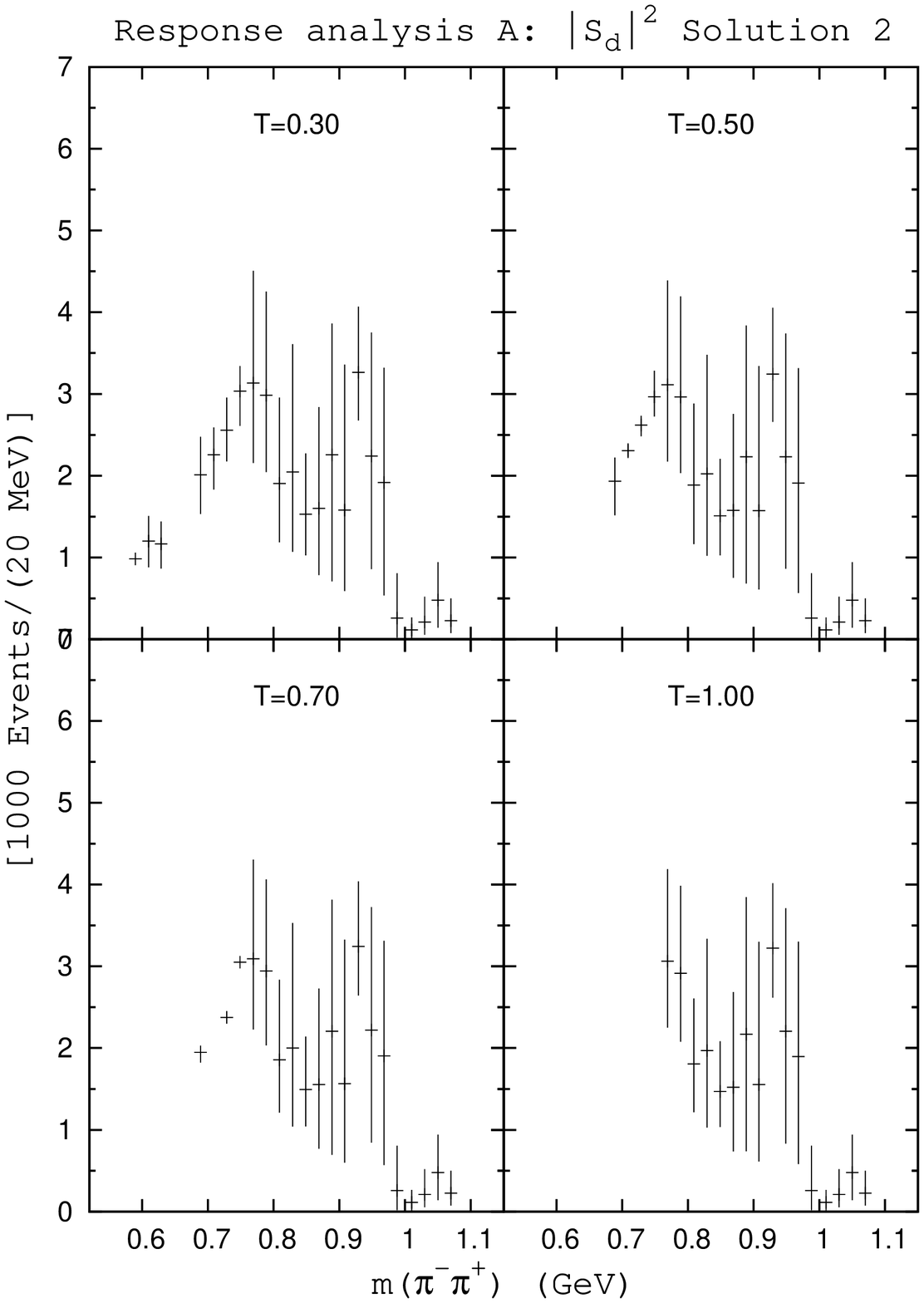}
\caption{Response of $S$-wave amplitude $|S_d|^2$ Solution 2 to
$|D^0_\tau|^2,|D^U_\tau|^2,|D^N_\tau|^2$ at high $T$.}
\label{Figure 2}
\end{figure}

\begin{figure} [htp]
\includegraphics[width=12cm,height=10.5cm]{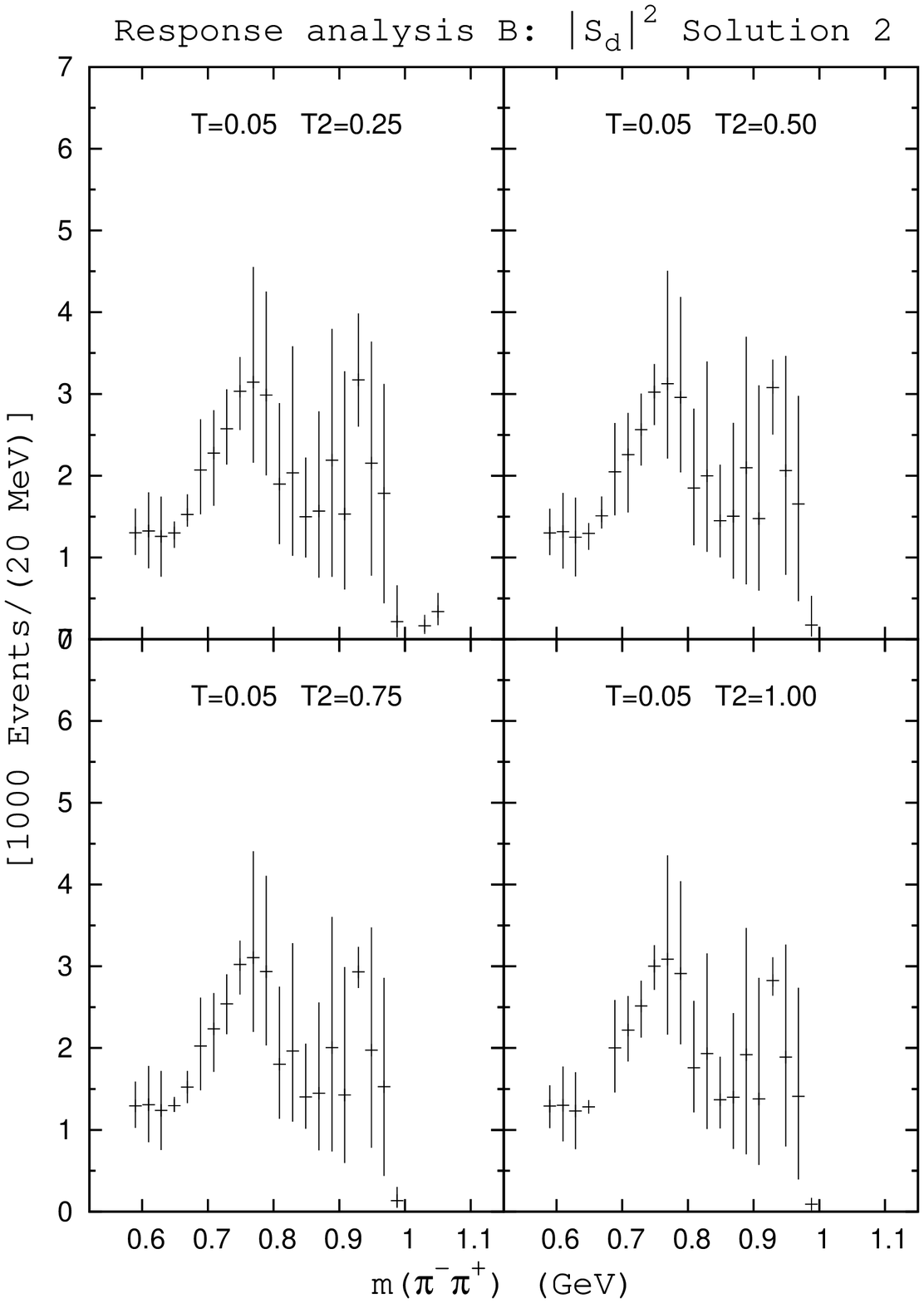}
\caption{Response of $S$-wave amplitude $|S_d|^2$ Solution 2 to
$|D^{2U}_\tau|^2,|D^{2N}_\tau|^2$ at $T=0.05$.}
\label{Figure 3}
\end{figure}

\begin{figure} [hp]
\includegraphics[width=12cm,height=10.5cm]{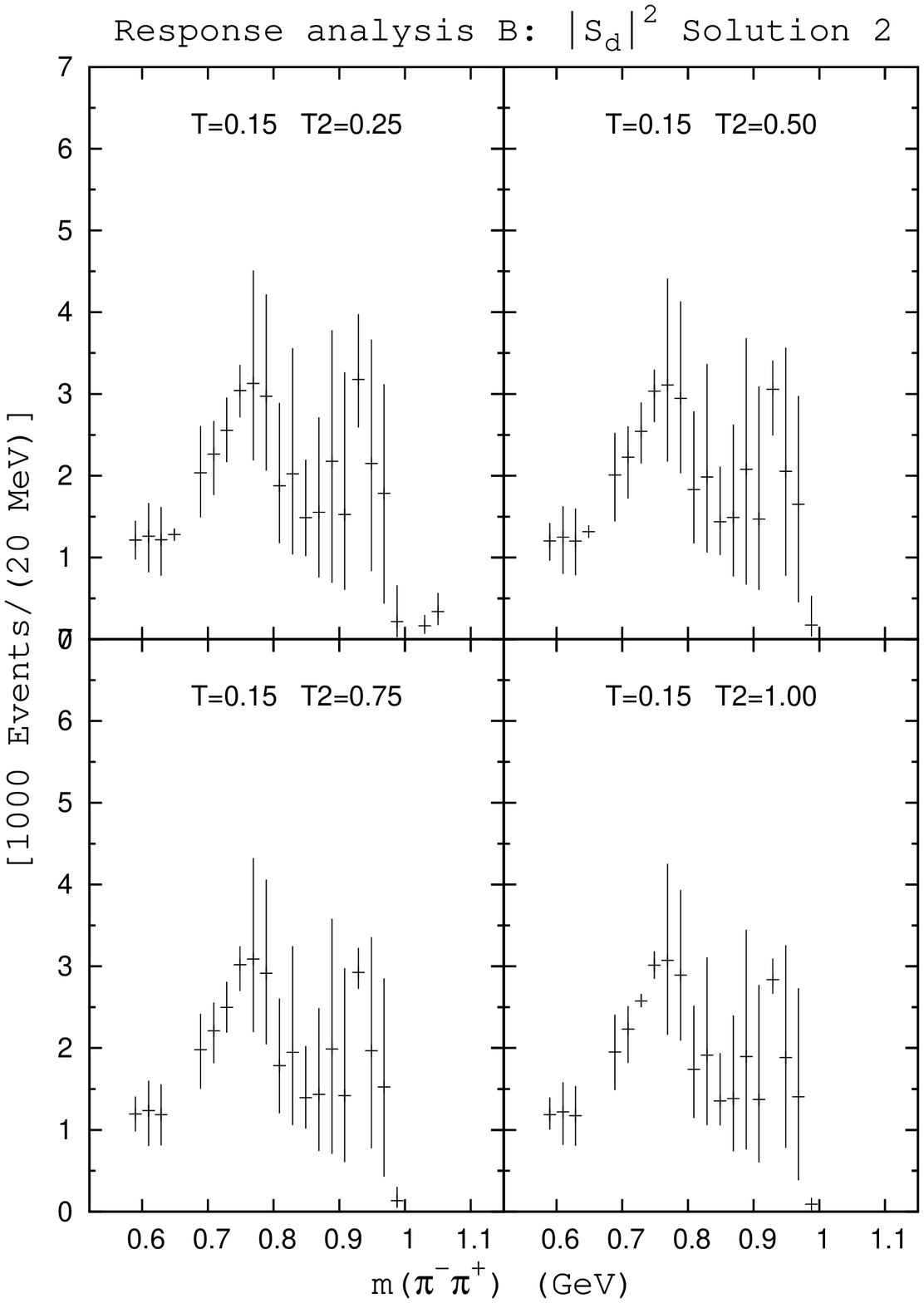}
\caption{Response of $S$-wave amplitude $|S_d|^2$ Solution 2 to
$|D^{2U}_\tau|^2,|D^{2N}_\tau|^2$ at $T=0.15$.}
\label{Figure 4}
\end{figure}

\begin{figure} [htp]
\includegraphics[width=12cm,height=10.5cm]{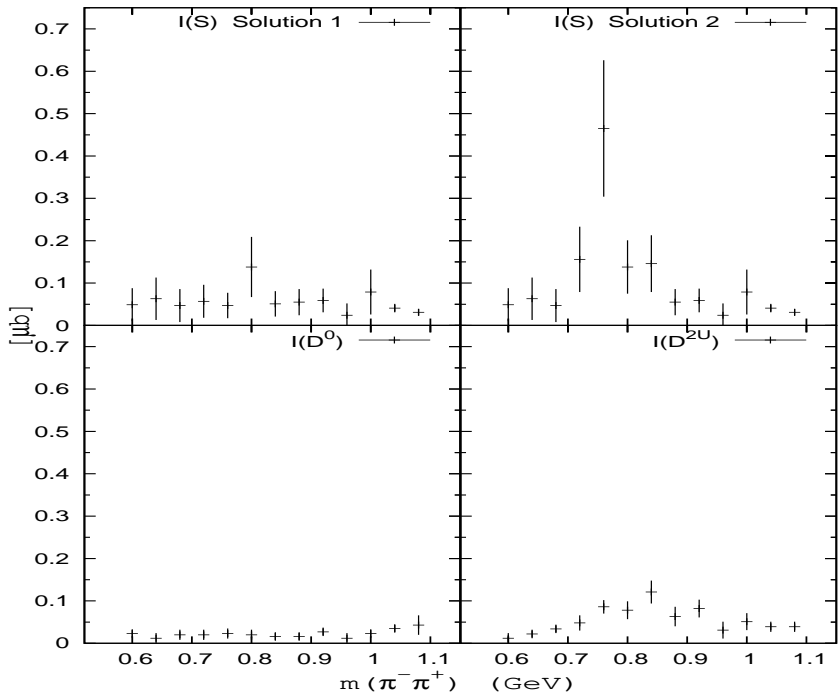}
\caption{$S$-wave and $D$-wave intensities from amplitude analysis at high $t$~\cite{rybicki85}.}
\label{Figure 5}
\end{figure}

Two solutions were found for the $S$- and $P$-wave amplitudes in the $\rho^0(770)$ mass region. The solutions for the $D$-wave amplitudes are unique for all dipion masses. 

Figure 5 shows the corresponding two solutions for the $S$-wave intensity $I(S)$ and $D$-wave intensities $I(D^0)$ and $I(D^{2U})$ below 1100 MeV. The other $D$-wave intensities are similar to $I(D^0)$. The Solution 2 of $I(S)$ suggests the presence of $\rho^0(770)$ in the $S$-wave indicating $\rho^0(770)-f_0(980)$ mixing at high $|t|$ even in the presence of a large $D$-wave amplitude $D^{2U}$. Unfortunately, some details of the $\rho^0(770)-f_0(980)$ mixing may be lost in both solutions for $I(S)$ since this analysis is done in a single broad bin $0.20 \leq |t| \leq 1.00$ (GeV/c$)^2$ and with 40 MeV mass bins. 

\section{Consistency of $\rho^0(770)-f_0(980)$ mixing with isospin relations between\\ $S$-wave amplitudes in $\pi^- p \to \pi^- \pi^+ n$ and 
$\pi^- p \to \pi^0 \pi^0 n$.}

\subsection{Isospin relations between $S$-wave intensities in $\pi^- \pi^+$, $\pi^0\pi^0$ and $\pi^+\pi^+$ production} 

If $\rho^0(770)-f_0(980)$ mixing is to be a genuine physical effect than it must be compatible with other $\pi N \to \pi\pi N$ processes whose amplitudes are related by isospin. Such isospin relations are the same for the $S$-matrix amplitudes and for the measured dephased amplitudes. An experimental test of these relations for $S$-wave amplitudes provides an independent evidence for or against the existence of the $\rho^0(770)-f_0(980)$ mixing in $\pi^- p \to \pi^- \pi^+ n$.

Generalized Bose-Einstein statistics is an extension of Bose-Einstein statistics to all particles belonging to an isospin multiplet of isospin $I_1$ which are regarded as $2I_1+1$ charge states of the same particle. By incorporating the isospin quantum numbers in the state vector the symmetrization properties are extended to the interchange of particles belonging to the same multiplet. The Generalized Bose-Einstein statistics requires that for two-pion state $J+I=even$ where $J$ is the dipion spin and $I$ is the total spin~\cite{martin70}.

Consider the two-pion state $|\pi^->|\pi^+>$. It can be written in the form
\begin{equation}
|\pi^-\pi^+>=\frac{1}{\sqrt{2}}|S>+\frac{1}{\sqrt{2}}|A>
\end{equation}
where $|S>$ and $|A>$ are symmetric and antisymmetric $\pi^-\pi^+$ isospin states
\begin{eqnarray}
|S> & = & \frac{1}{\sqrt{2}}(|\pi^->| \pi^+>+|\pi^+>|\pi^->)=
-\frac{1}{\sqrt{3}}(\sqrt{2}|0,0>+|2,0>)\\
\nonumber
|A> & = & \frac{1}{\sqrt{2}}(|\pi^->| \pi^+>-|\pi^+>|\pi^->) = |1,0>
\end{eqnarray}
where we used the convention $|\pi^+>=-|1,+1>$~\cite{gibson76}. In $\pi^-\pi^+$ channel the $J=even$ and $J=odd$ transversity amplitudes thus involve isospin states $\frac{1}{\sqrt{2}}|S>$ and $\frac{1}{\sqrt{2}}|A>$, respectively. The two-pion states $|\pi^0 \pi^0>$ and $|\pi^+ \pi^+>$  correspond to even $I$
\begin{eqnarray}
|\pi^0 \pi^0> & = & -\frac{1}{\sqrt{3}}(|0,0>-\sqrt{2}|2,0>)\\
\nonumber
|\pi^+\pi^+> & = & |2,2>
\end{eqnarray}
which implies only $J=even$ dipion states are allowed..

We now consider $S$-wave transversity amplitudes $S_\tau(-+)$, $S_\tau(00)$ and $S_\tau(++)$ in three processes $\pi^- p \to \pi^- \pi^+ n$, $\pi^- p \to \pi^0 \pi^0 n$ and $\pi^+ p \to \pi^+ \pi^+ n$. Using (5.2) and (5.3) we can express these amplitudes in terms of amplitudes $S_\tau^{II_3}$ with definite total isospin $I$ and $I_3$

\begin{subequations}
 \label{allequations} 
 \begin{eqnarray}
  S_\tau(-+)&=&-{1 \over{\sqrt{3}}} \{ S^{00}_\tau + {1 \over{2}}
  \sqrt{2}    S^{20}_\tau \} \label{equationa}
  \\
  S_\tau(00)&=&-{1 \over{\sqrt{3}}} \{ S^{00}_\tau - \sqrt{2}
  S^{20}_\tau\} \label{equationb}
  \\
  S_\tau(++)&=&S^{22}_\tau \label{equationc}
 \end{eqnarray}
\end{subequations}
Assuming the invariance of the amplitudes  $S_\tau(c)$, $c=(-+),(00),(++)$ under the rotations in isospin space, the isospin amplitudes $S^{II_3}_\tau$ then do not depend on the  component $I_3$ and we have
\begin{equation}
S^{20}_\tau = S^{22}_\tau = S_\tau(++)
\end{equation}
From (5.4a) and (5.4b) we then get the isospin relation between the $S$-wave amplitudes
\begin{equation}
S_\tau(00)=S_\tau(-+)+\sqrt{{3 \over{2}}} S_\tau(++)
\end{equation}
It is useful to write the following combinations of this equation
\begin{subequations}
 \label{allequations} 
 \begin{eqnarray}
  \sqrt{{3 \over{2}}} S_\tau(++)&=&S_\tau(00)-S_\tau(-+)
  \\
  S_\tau(-+)&=&S_\tau(00) - \sqrt{{3 \over{2}}} S_\tau(++)
  \\
  S_\tau(00)&=&S_\tau(-+) + \sqrt{{3 \over{2}}} S_\tau(++)
 \end{eqnarray}
\end{subequations}
and calculate the $S$-wave intensities
\begin{equation}
I_S(c)=|S_u(c)|^2 + |S_d(c)|^2 
\end{equation}
for $c=(++),(-+),(00)$ using expressions on r.h.s. of (5.7). With
\begin{equation}
I_S(2)={3 \over{2}}I_S(++)
\end{equation}
we then obtain
\begin{subequations}
 \label{allequations} 
 \begin{eqnarray}
  I_S(-+)+I_S(00) - I_S(2)&=&+2 \sum\limits_\tau
  Re[S_\tau(-+)S^*_\tau(00)] \\
  I_S(-+)-I_S(00) - I_S(2)&=&-2 \sqrt{{3 \over {2}}} \sum\limits_\tau
  Re[S_\tau(00)S^*_\tau(++)] \\
  I_S(-+)-I_S(00) + I_S(2)&=&-2 \sqrt{{3 \over {2}}} \sum\limits_\tau
  Re[S_\tau(-+)S^*_\tau(++)]
 \end{eqnarray}
\end{subequations}

\begin{figure} [htp]
\includegraphics[width=12cm,height=10cm]{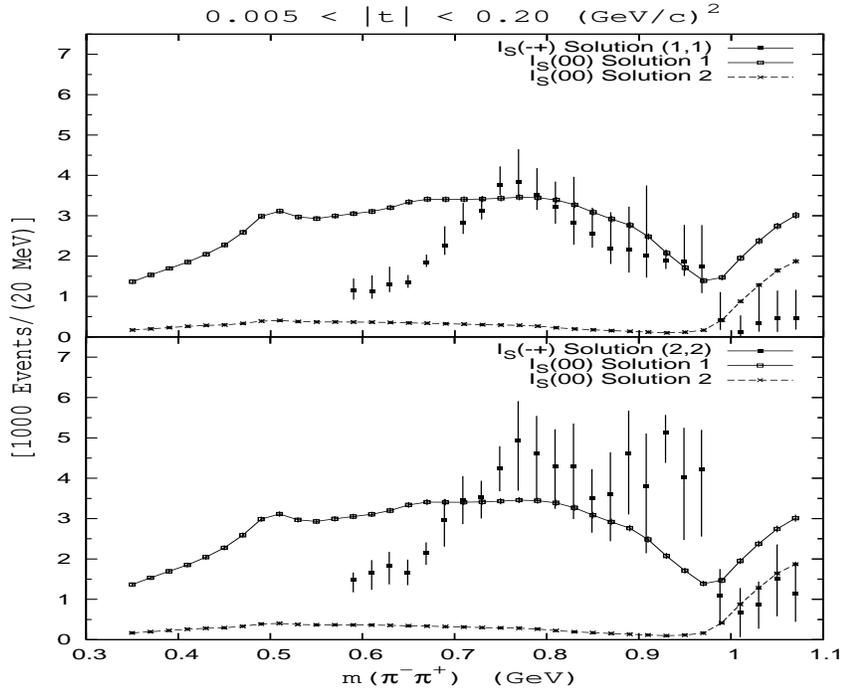}
\caption{Comparison of intensity $I_S(-+)$ from Ref.~\cite{svec12b} with the intensity $I_S(00)$ from Ref.~\cite{gunter01}.}
\label{fig6}
\end{figure}

The interference terms on the r.h.s. of (5.10) are scalar products $<A_S(c)| A_S(c')>$ of four-vectors
\begin{equation}
|A_S(c)>=(Re S_u(c), Im S_u(c), Re S_d(c), Im S_d(c))
\end{equation}
in a Euclidian 4-dimensional space with the norm $<A_S(c)|A_S(c)>=I_S(c)$ and scalar product
\begin{equation}
<A_S(c)| A_S(c')> = \sqrt{I_S(c)} \sqrt{I_S(c')} \cos \Omega_{cc'}(S)
\end{equation}

\begin{figure} [htp]
\includegraphics[width=12cm,height=10cm]{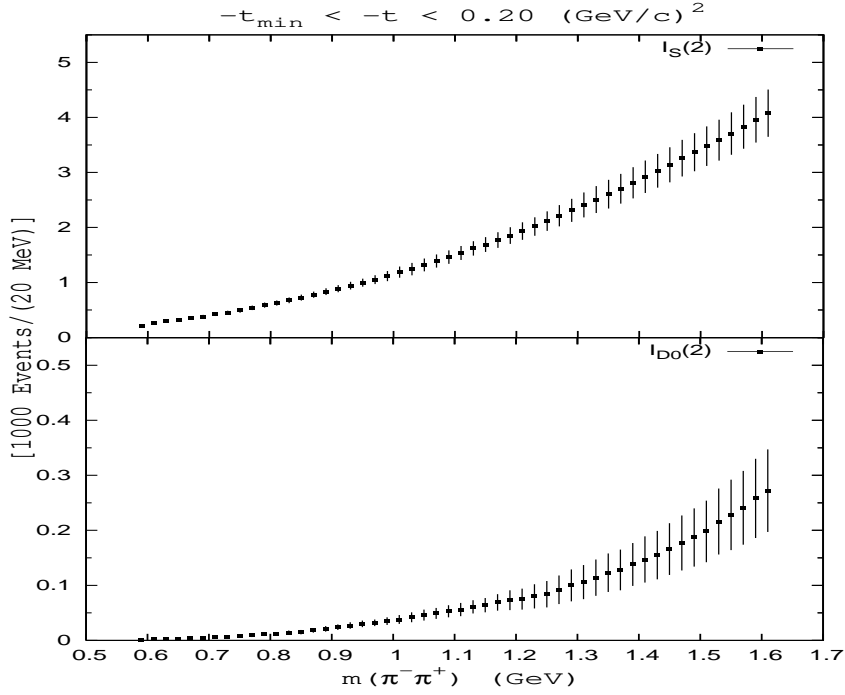}
\caption{Intensities $I_S(2)$ and $I_{D^0}(2)$ from CERN data on $\pi^+ \pi^+$ at 12.5 GeV/c~\cite{hoogland77} scaled to 17.2 GeV/c.}
\label{fig7}
\end{figure}

where $\Omega_{cc'}(S)$ is an angle between the vectors $A_S(c)$ and $A_S(c')$. The relations (5.10) then read
\begin{subequations}
 \label{allequations} 
 \begin{eqnarray}
  I_S(-+)+I_S(00) - I_S(2)&=&+2 \sqrt{I_S(-+)} \sqrt{I_S(00)} \cos
  \Omega_1(S)
  \\
  I_S(-+)-I_S(00) - I_S(2)&=&-2 \sqrt{I_S(00)} \sqrt{I_S(2)} \cos
  \Omega_2(S)
  \\
  I_S(-+)-I_S(00) + I_S(2)&=&-2 \sqrt{I_S(-+)} \sqrt{I_S(2)} \cos
  \Omega_3(S)
 \end{eqnarray}
\end{subequations}
The equations (5.13) represent three linearly independent constraints on the measured spectra $I_S(-+)$, $I_S(00)$ and $I_S(++)= {2 \over{3}}I_S(2)$ imposed by the requirement that the cosines have physical values. While the cosines are linearly independent, they satisfy a non-linear constraint
\begin{equation}
\cos^2 \Omega_1(S) + \cos^2 \Omega_2(S) + \cos^2 \Omega_3(S) -2 \cos \Omega_1(S) \cos \Omega_2(S) \cos \Omega_3(S) = 1
\end{equation}
The constraint (5.14) implies that for physical values of the cosines the phases satisfy a phase condition
\begin{equation}
\Omega_1(S)+\Omega_2(S)+\Omega_3(S)=0
\end{equation}

\subsection{Data used in the test of isospin relations for the $S$-waves}

To test the constraints (5.13) we need data on the intensities $I_S(c)$ from $\pi^- \pi^+$, $\pi^0 \pi^0$ and $\pi^+ \pi^+$ production in the dipion mass interval $580 \leq m \leq 1080$ MeV where we observe $\rho^0(770)-f_0(980)$ mixing. The available data allow to perform the tests at small momentum transfers $0.005 < |t| < 0.20$ (GeV/c)$^2$.

For the $\pi^- \pi^+$ channel we used our high resolution analysis using Monte Carlo method presented in Ref.~\cite{svec12b}. It produced two solutions for the moduli $|S_u(i)|^2,i=1,2$ and $|S_d(j)|^2,j=1,2$. Two solutions for the $S$-wave intensity $I_S(-+)$ were used in this mass range corresponding to combinations (1,1) and (2,2) of solutions for the moduli
\begin{eqnarray}
I_S(-+) \text{ Solution (1,1)} & = & |S_u(1)|^2 +|S_d(1)|^2\\
\nonumber
I_S(-+) \text{ Solution (2,2)} & = & |S_u(2)|^2 +|S_d(2)|^2
\end{eqnarray}
The results for $I_S(-+)$ from our analysis are shown in Figure 6. The unit for $d^2\sigma/dtdm$ in Ref.~\cite{grayer74} used in our analysis can be converted to $\mu b/20$ MeV using a conversion factor $0.109 \mu b/20$ MeV = 1000 events/20 MeV.

For the $\pi^0 \pi^0$ channel we used the BNL data at 18.3 GeV/c~\cite{gunter01}. The BNL data were converted from native BNL units "intensity/40 MeV" into our units "1000 events/20 MeV" using a conversion factor $F=0.6700\times10^{-4}$. We obtained this factor by comparing the $f_2(1270)$ peak value in their Figure 5F given in units "intensity/40 MeV" with the value of coresponding 4 bins at $f_2(1270)$ peak in their Figure 4a given in units "events/10 MeV". The data in two bins $0.01 < |t| < 0.10$ (GeV/c)$^2$ and $0.10 < |t| < 0.20$ (GeV/c)$^2$ were combined by addition to a sigle bin $0.01 < |t| < 0.20$ (GeV/c)$^2$ corresponding to the CERN measurements. The data were then interpolated to 20 MeV bins and scaled to 17.2 GeV/c using phase and flux factor $K(s,m^2)$ given by~\cite{svec97a}
\begin{eqnarray}
K(s,m^2) & = & {G(s,m^2) \over {\text{Flux}(s)}}\\
\nonumber
G(s,m^2) & = & {1 \over {(4 \pi )^3}}{q(m^2) \over {\sqrt{[s-(M+\mu)^2][s-(M-\mu)^2]}}}\\
\nonumber
\text{Flux}(s) & = & 4Mp_{\pi lab}
\end{eqnarray}

where $q(m^2)={1 \over {2}} \sqrt{m^2 -4\mu^2}$ is the pion momentum in the center of mass of the dipion system of mass $m$, and $M$ and $\mu$ are the nucleon and pion mass, respectively. The two Solutions 1 and 2 for intensities $I_S(00)$ are shown in Figure 6 and compared with the corresponding intensities in $\pi^- p \to \pi^- \pi^+ n$.

For the $\pi^+ \pi^+$ channel we used CERN data on $\pi^+ p \to \pi^+ \pi^+ n$ at 12.5 GeV/c~\cite{hoogland77}. This data was used to determine the isospin $I$=2 $S$- and $D$-wave amplitudes $f^{I=2}_{J=0}$ and $f^{I=2}_{J=2}$ in $\pi \pi$ scattering using a pion exchange formula for $S$- and $D^0$-wave helicity flip amplitudes $A_1(++)$, $A=S,D^0$
\begin{equation}
A_1(++)=F(s,t,m^2) \sqrt{2J+1}  f^{I=2}_J
\end{equation}
with an assumption that the non-flip amplitudes vanish $A_0(++)$=0. The partial wave intensities $I_S(++)$ and $I_{D^0}(++)$ at 17.2 GeV/c were then reconstructed using
\begin{equation}
I_A(++)=|A_1(++)|^2 \text{  and  } I_A(2)={3 \over {2}} I_A(++)
\end{equation}
The common factor $|F(s,t,m^2)|^2$ was taken from the analysis of Kaminski {\sl et al} ~\cite{kaminski02} by identifying the mean values of our $I_S(2)$ with the values of their $S$-wave $I=2$ contribution $I(2)$ at 17.2 GeV/c and $0.005<|t|<0.20$ (GeV/c)$^2$ presented in their Figure 2. The errors on $I_A(2)$ are given by the errors on $f^{I=2}_{J=0,2}$ and are taken from the CERN analysis in Ref.~\cite{hoogland77}. The results for $I_S(2)$ and $I_{D^0}(2)$ were scaled to 17.2 GeV/c and are shown in Figure 7.

\begin{figure} [htp]
\includegraphics[width=12cm,height=10cm]{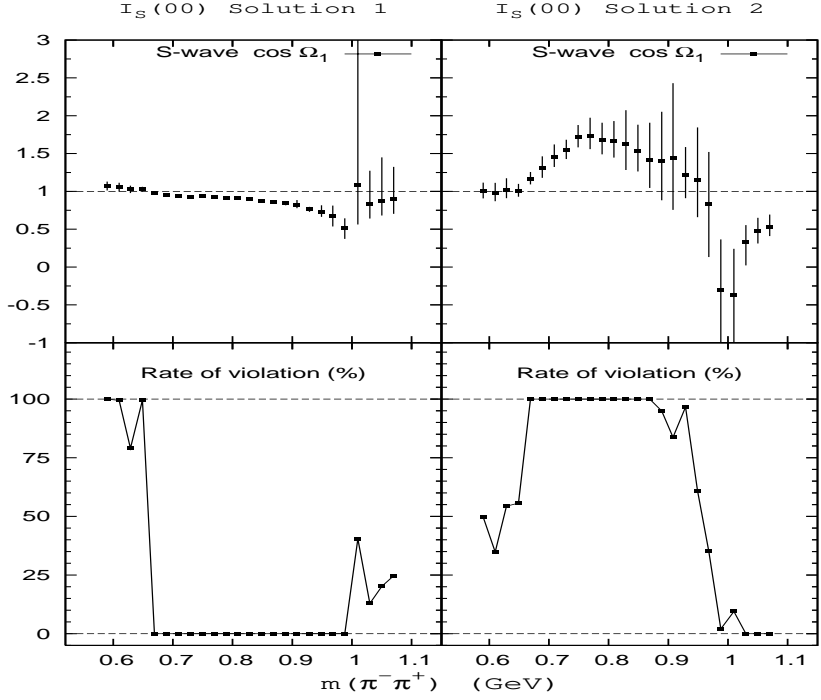}
\caption{Test of isospin relations with Solution (1,1) for $S$-wave intensity $I_S(-+)$ from analysis~\cite{svec12b}.}
\label{fig8}
\end{figure}

\begin{figure} [hp]
\includegraphics[width=12cm,height=10cm]{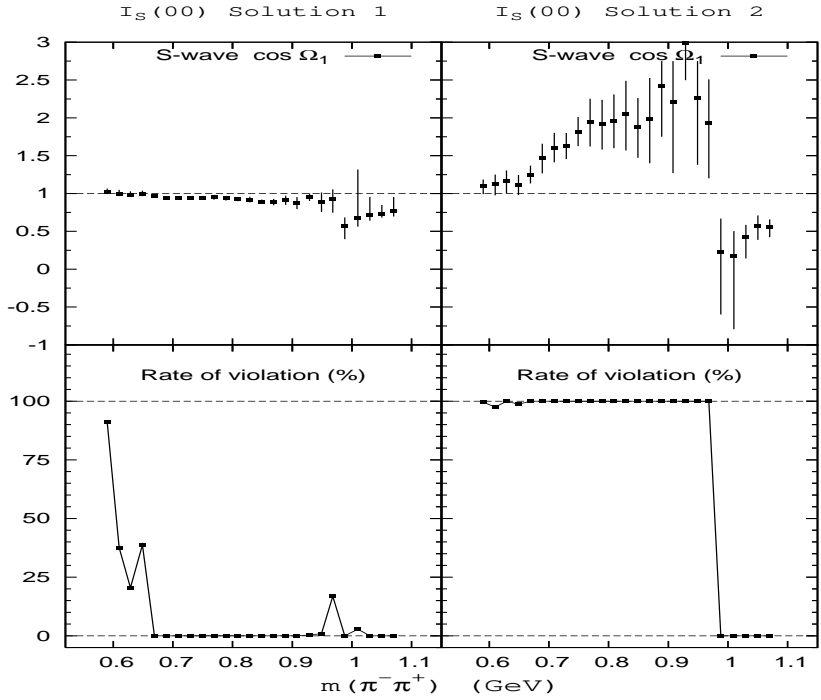}
\caption{Test of isospin relations with Solution (2,2) for $S$-wave intensity $I_S(-+)$ from analysis~\cite{svec12b}.}
\label{fig9}
\end{figure}

\subsection{Data analysis and the results}

The three isospin constraints (5.13) were tested as follows. The error volume of the intensities $I_S(-+)$, $I_S(00)$ and $I_S(2)$ was sampled by Monte Carlo method in each mass bin. For each of the 10,000 data samplings the equations (5.13) were used to calculate the cosines $\cos \Omega_1(S)$, $\cos \Omega_2(S)$ and $\cos \Omega_3(S)$ for all solution combinations of the $S$-wave amplitudes. A distribution of values of $\cos \Omega_k(S)$ has been obtained for each $k=1,3$ which defined the range and average value of $\cos \Omega_k(S$ in each mass bin. In general, these average values of $\cos \Omega_k(S)^{av}$ were close to the mean values of $\cos \Omega_k(S)^*$ calculated from the mean values of the intensities.

In each mass bin a number count was taken of the physical and unphysical values of $\cos \Omega_k(S)$ to quantify any possible violation of the constraints (5.13). The program also verified that the non-linear condition (5.14) on the cosines is satisfied for both the physical and unphysical values of $\cos \Omega_k(s)$, $k=1,2,3$ for each Monte Carlo sampling. Importantly, the number counts for unphysical value of $\cos \Omega_k(S)$ were identical for all three cosines in each mass bin. In the Figures 8 and 9 we thus present only the results for $\cos \Omega_1(S)$ and the corresponding fraction of unphysical values of $\cos \Omega_1(S)$ to evaluate the degree of violation of constraints (5.13).

Figures 8 and 9 show the results for the Solutions (1,1) and (2,2) of $I_S(-+)$, respectively. For Solution 1 of $I_S(00)$ and for both solutions of $I_S(-+)$ the isospin relations are clearly satisfied although there is a small violation of the constraints (5.13) below 680 MeV. For Solution 2 of $I_S(00)$ there is a massive violation of constraints (5.13) in the mass range 580-980 MeV for both solutions of $I_S(-+)$. This suggests that this small-valued solution for $I_S(00)$ is unphysical.

\section{Survey of evidence for $\rho^0(770)-f_0(980)$ spin mixing from amplitude analyses on polarized targets.}

There are five measurements of production $\pi N \to \pi^-\pi^+ N$ on polarized target at four energies:\\
(1) CERN measurement of $\pi^-p \to \pi^-\pi^+n$ at 17.2 GeV/c at $0.005 < |t| < 0.20$ (GeV/c)$^2$~\cite{becker79a,becker79b,chabaud83}\\
(2) CERN measurement of $\pi^-p \to \pi^-\pi^+n$ at 17.2 GeV/c at $0.20 < |t| < 1.00$ (GeV/c)$^2$~\cite{rybicki85}\\
(3) ITEP measurement of $\pi^-p \to \pi^-\pi^+n$ at 1.78 GeV/c at $0.005 < |t| < 0.20$ (GeV/c)$^2$~\cite{alekseev98,alekseev99}\\
(4) CERN-Saclay measurement of $\pi^+ n \to \pi^+ \pi^- p$ at 5.98 GeV/c at $0.20 < |t| < 0.40$ (GeV/c)$^2$~\cite{lesquen85}\\
(5) CERN-Saclay measurement of $\pi^+ n \to \pi^+ \pi^- p$ at 11.85 GeV/c at $0.20 < |t| < 0.40$ (GeV/c)$^2$~\cite{lesquen85}

There are several amplitude analyses of the CERN measurement at low $t$~\cite{becker79a,becker79b,chabaud83,svec96,svec97a} with the most recent being Ref.~\cite{kaminski97,svec12b,svec14a}. Amplitude analyses of the CERN measurement at high $t$ and of the ITEP measurement are presented in Ref.~\cite{rybicki85} and Ref.~\cite{alekseev98,alekseev99}, respectively. Amplitude analyses of the CERN-Saclay (CS) data are presented in Ref.~\cite{svec92a,svec96,svec97a} and in this work (Figures 12 and 13). The analyses used different methods to determine the amplitudes and their errors: $\chi^2$ fits~\cite{becker79a,becker79b,chabaud83,rybicki85,kaminski97}, Monte Carlo analytical solutions~\cite{svec96,svec97a,svec12b,svec14a} and analytical solutions with error propagation~\cite{svec92a,alekseev98,alekseev99}. All amplitude analyses are mutually consistent in providing a clear evidence for $\rho^0(770)-f_0(980)$ mixing in the $S$-wave transversity amplitudes. 

\subsection{Amplitude analyses of the ITEP and CERN-Saclay data}

Figure 10 shows the $S$-wave intensity $I(S)$ and $P$-wave intensity $I(L)$ from the analysis of the ITEP data~\cite{alekseev98,alekseev99} at low t. Both Solutions for $I(S)$ indicate evidence for $\rho^0(770)$ structure at 770 MeV but this structure is clearly prominent in the Solution 2. The moduli of $S$-wave transversity amplitudes are shown in Figure 11 with a clear $\rho^0(770)$ signal in the Solution 2. Note that in this analysis $I(S) \approx I(L)$ in the Solutions 2 at $\rho^0(770)$ mass. The ITEP analysis is particularly interesting in that it provides a support for the 1960's low energy analyses which first indicated a rho-like structure in the $S$-wave~\cite{hagopian63,islam64,patil64,durand65,baton65}. A review of these early analyses is given in~\cite{gasiorowicz66}.

Figure 12 shows $S$-wave intensity from our Monte Carlo amplitude analysis of the CERN-Saclay data on $\pi^+ n \to \pi^+ \pi^- p$ at 5.58 and 11.85 GeV/c at    $0.20\leq |t| \leq 0.40$ (GeV/c)$^2$. At both energies the Solution (1,1) is smaller than the Solution (2,2) which clearly resonates at $\rho^0(770)$ mass.

\begin{figure} [htp]
\includegraphics[width=12cm,height=10.5cm]{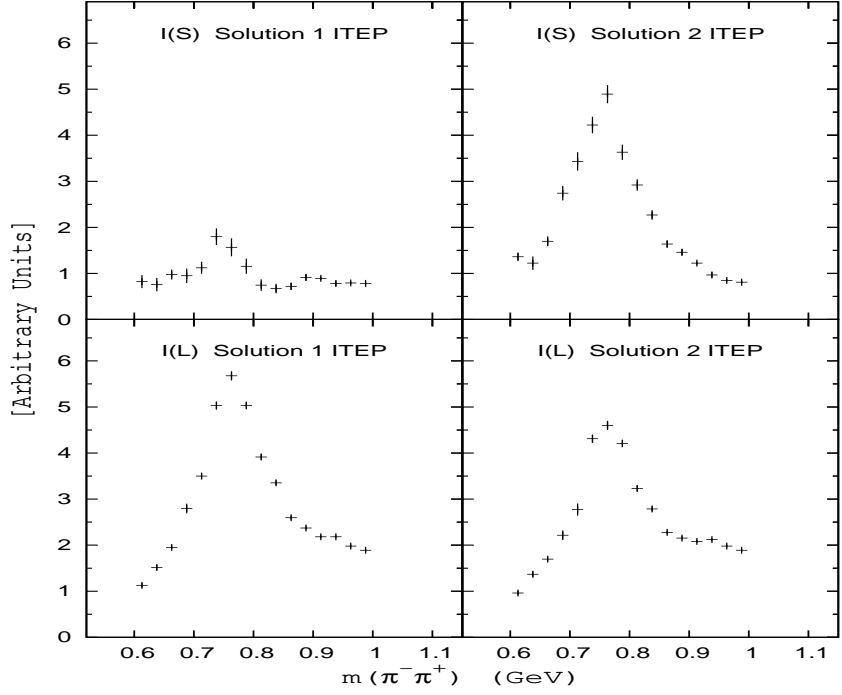}
\caption{Intensities $I(S)$ and $I(L)$ in $\pi^- p \to \pi^- \pi^+ n$ at 1.78 GeV/c at low $t$. Data from Ref~\cite{alekseev99}.}
\label{Figure 10}
\end{figure}

\begin{figure} [hp]
\includegraphics[width=12cm,height=10.5cm]{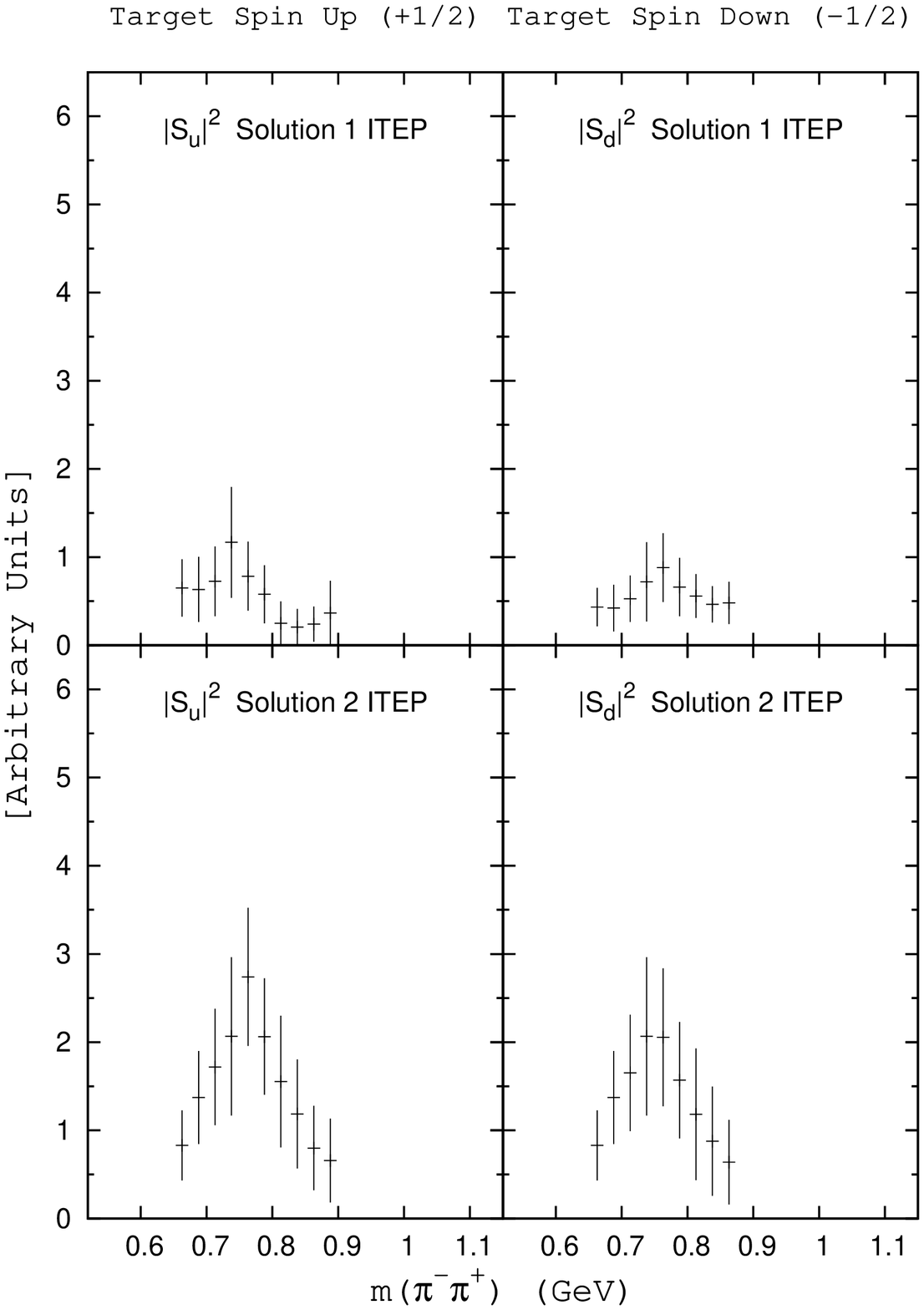}
\caption{$S$-wave moduli $|S_\tau|^2$ in $\pi^- p \to \pi^- \pi^+ n$ at 1.78 GeV/c at low $t$. Data from Ref~\cite{alekseev99}.}
\label{Figure 11}
\end{figure}

\begin{figure} [htp]
\includegraphics[width=12cm,height=10.5cm]{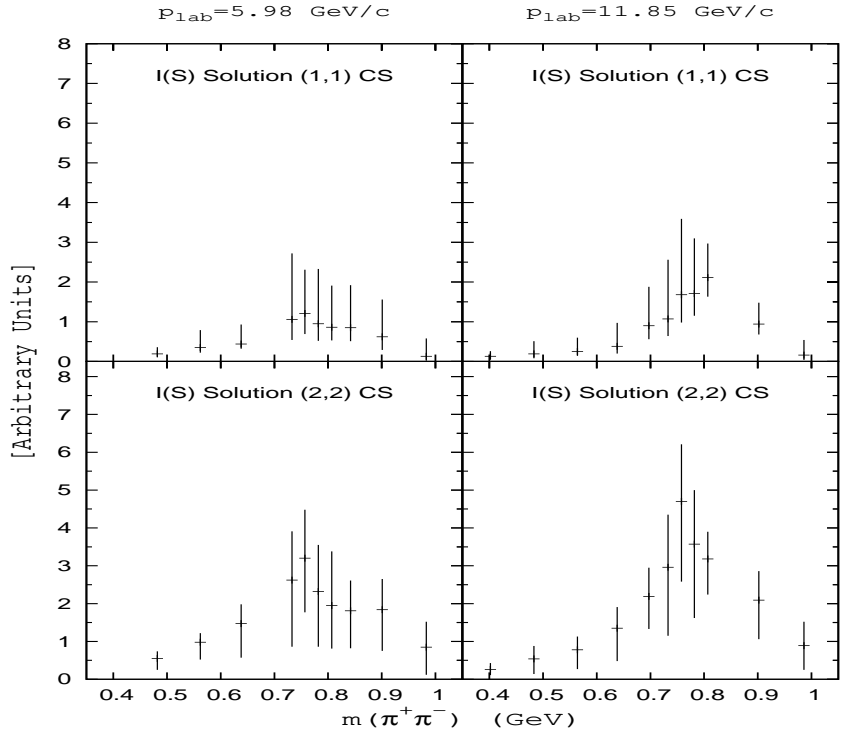}
\caption{Intensities $I(S)$ in $\pi^+ n \to \pi^+ \pi^- n$ at 5.98 and 11.85 GeV/c at high $t$. Data from Ref~\cite{lesquen85}.}
\label{Figure 12}
\end{figure}

\begin{figure} [hp]
\includegraphics[width=12cm,height=10.5cm]{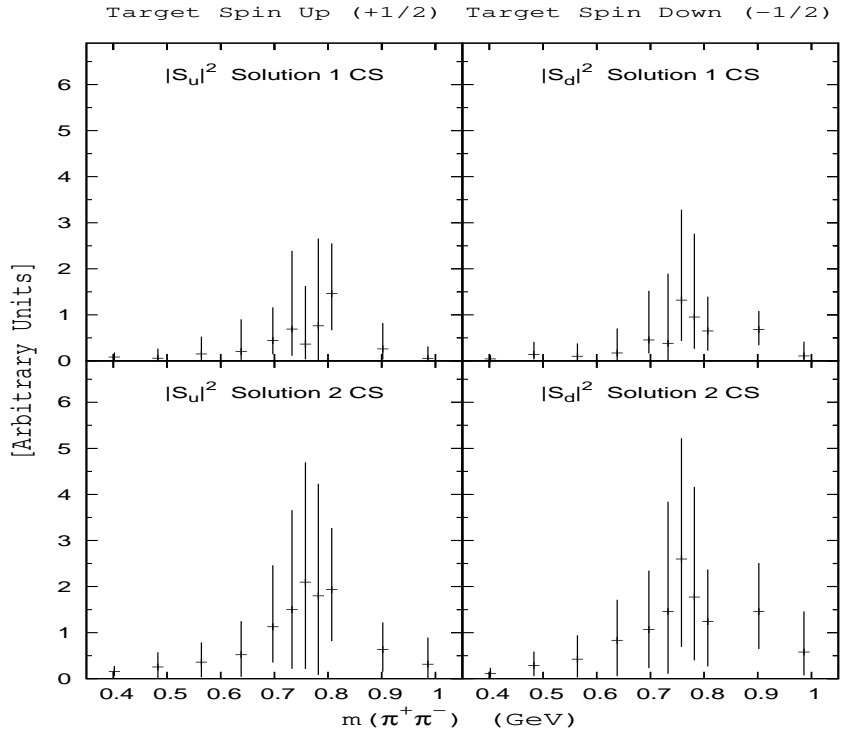}
\caption{$S$-wave moduli $|S_\tau|^2$ in $\pi^+ n \to \pi^+ \pi^- n$ at 11.85 GeV/c at high $t$. Data from Ref~\cite{lesquen85}.}
\label{Figure 13}
\end{figure}

The corresponding $S$-wave moduli at 11.85 GeV/c are shown in Figure 13. At these larger $t$ the Solution 1 is small for both transversity amplitudes while the Solution 2 shows a clear $\rho^0(770)$ structure in both $|S_u|^2$ and 
$|S_d|^2$. The results for the amplitudes at 5.98 GeV/c are very similar~\cite{svec96,svec97a}. These results with one million Monte Carlo sampling of the data error volume are very similar to our analysis of the same data with a smaller sampling rate in Ref.~\cite{svec96}.

\subsection{Amplitude analyses of the CERN-Cracow-Munich data}

Figure 14 shows the $S$-wave intensity from the amplitude analysis of the CERN data at high $t$~\cite{rybicki85}. Again at these larger $t$ the Solution 1 is small while the Solution 2 suggests a pronounced $\rho^0(770)$ structure.

Authors of Ref.~\cite{becker79a}(referred to as $\chi^2$ 79) present normalized moduli $|\tilde{S}_\tau|^2$. Figure 15 presents the corresponding unnormalized moduli $|S_\tau|^2=|\tilde{S}_\tau|^2 d^2\sigma /dtdm$ in 20 MeV bins from 600-900 MeV at $0.005 < |t| < 0.20$ (GeV/c)$^2$. Authors of Ref.~\cite{becker79b,kaminski97} present $S$-wave intensity $I(S)=|S_u|^2+|S_d|^2$ and the ratio of the moduli $R=|S_u|/|S_d|$. From this data it is a simple matter to reconstruct the moduli $|S_\tau|^2$. Errors were calculated using the formalism for error propagation~\cite{taylor97}. The moduli from the analysis~\cite{becker79b} in 40 MeV bins from 600-1520 MeV at $0.010 < |t| < 0.20$ (GeV/c)$^2$ are presented in Figure 5 of Ref.~\cite{svec97a} and are not reproduced here. These results agree with our original analysis~\cite{svec97a} and with our new Analysis I~\cite{svec12b}. The moduli from the 1997 analysis by Kami\'{n}ski, Le\'{s}niak and Rybicki~\cite{kaminski97}(referred to as $\chi^2$ 97) in 20 MeV bins from 600-1600 MeV at $0.005 < |t| < 0.20$ (GeV/c)$^2$ are shown in Figure 16 below 1080 MeV. Figure 17 presents our results for the moduli from Analysis I on polarized target~\cite{svec12b} using the same input data~\cite{rybicki96} as the analysis $\chi^2$ 97 in the the Figure 16. The authors of amplitude  analyses~\cite{chabaud83,rybicki85} present only results for partial wave intensities.

Figures 15, 16 and 17 from the three amplitude  analyses~\cite{becker79a,kaminski97,svec12b} and the Figure 5 in Ref.~\cite{svec97a} from the amplitude analysis~\cite{becker79b} show a clear presence of a $\rho^0(770)$ structure in both Solutions for the amplitude $|S_d|^2$. Our Monte Carlo Analysis I in Figure 17 and the $\chi^2$ 97 analysis of Kami\'{n}ski, Le\'{s}niak and Rybicki in Figure 16 are nearly identical. This is not suprising since both analyses use the same data set~\cite{rybicki96} and their methods of amplitude analysis are both legitimate methods. Both analyses thus provide identical evidence for $\rho^0(770)-f_0(980)$ mixing. 

\begin{figure} [htp]
\includegraphics[width=12cm,height=10.5cm]{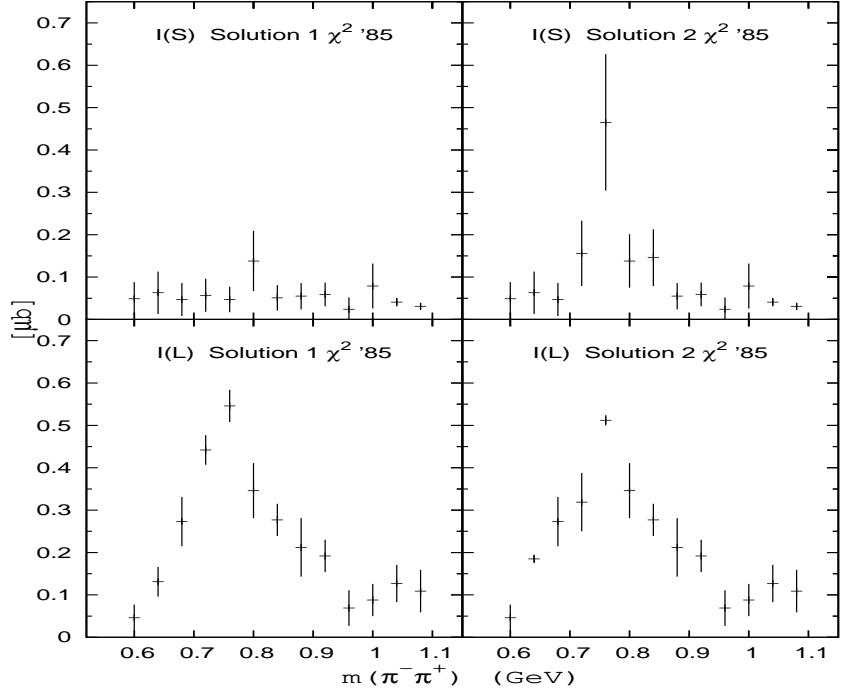}
\caption{Intensities $I(S)$ and $I(L)$ in $\pi^- p \to \pi^- \pi^+ n$ at 17.2 GeV/c at high $t$. Data from Ref~\cite{rybicki85}.}
\label{Figure 14}
\end{figure}

\begin{figure} [hp]
\includegraphics[width=12cm,height=10.5cm]{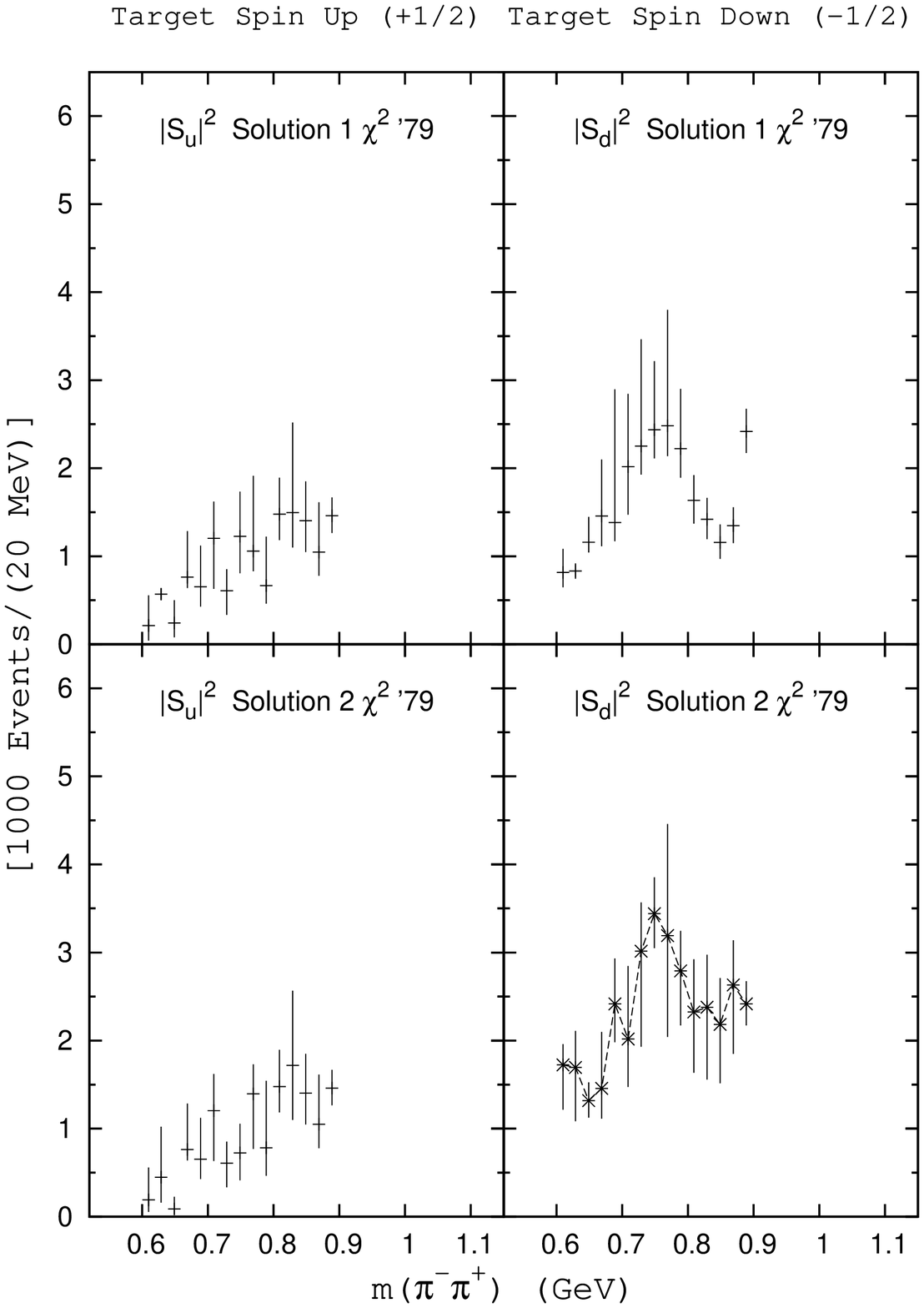}
\caption{$S$-wave moduli $|S_\tau|^2$ in $\pi^- p \to \pi^- \pi^+ n$ at 17.2 GeV/c at low $t$. Data from Ref.~\cite{becker79a}.}
\label{Figure 15}
\end{figure}

\begin{figure} [htp]
\includegraphics[width=12cm,height=10.5cm]{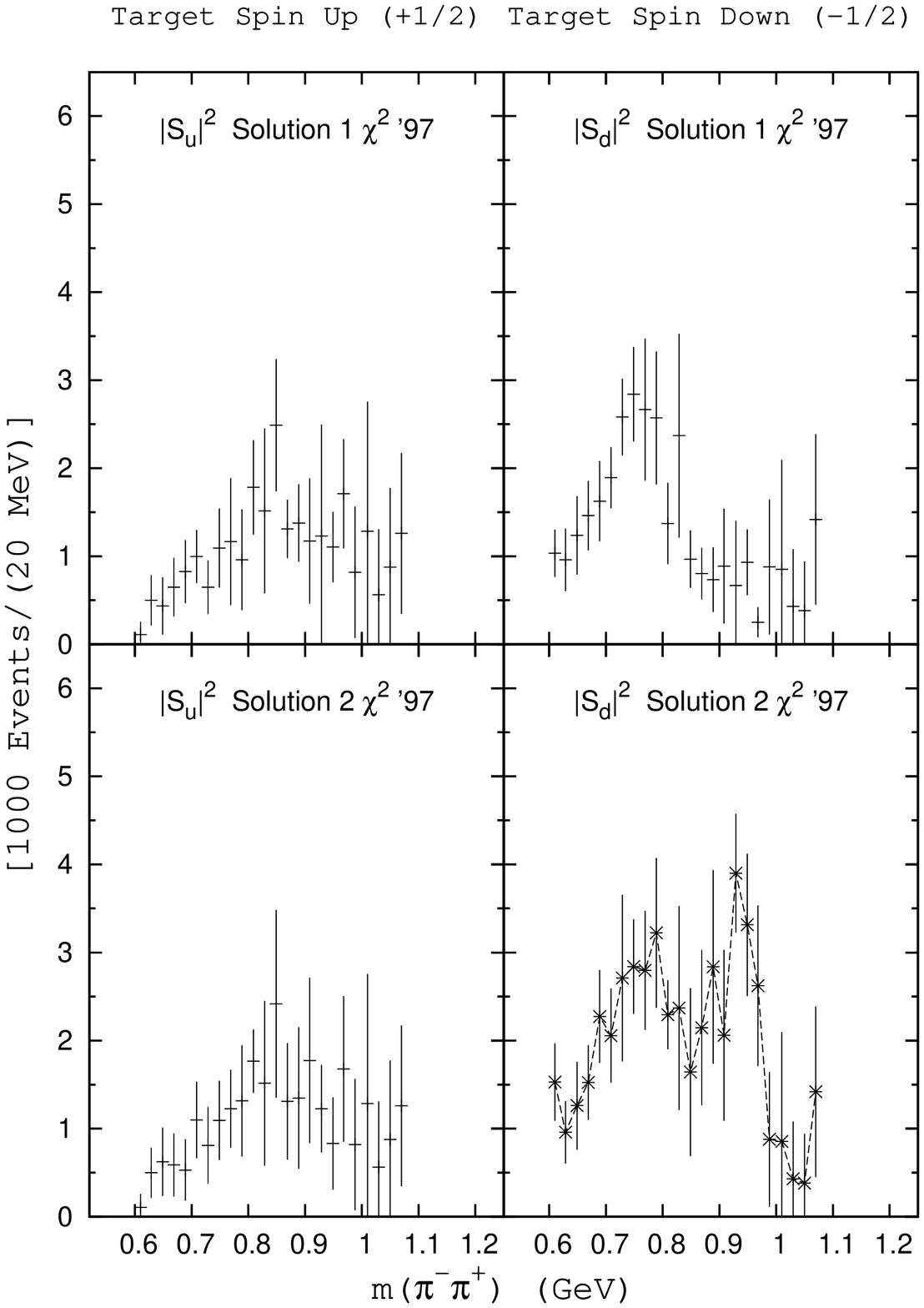}
\caption{$S$-wave moduli $|S_\tau|^2$ in $\pi^- p \to \pi^- \pi^+ n$ at 17.2 GeV/c at low $t$. Data from Ref~\cite{kaminski97}.}
\label{Figure 16}
\end{figure}

\begin{figure}[hp]
\includegraphics[width=12cm,height=10.5cm]{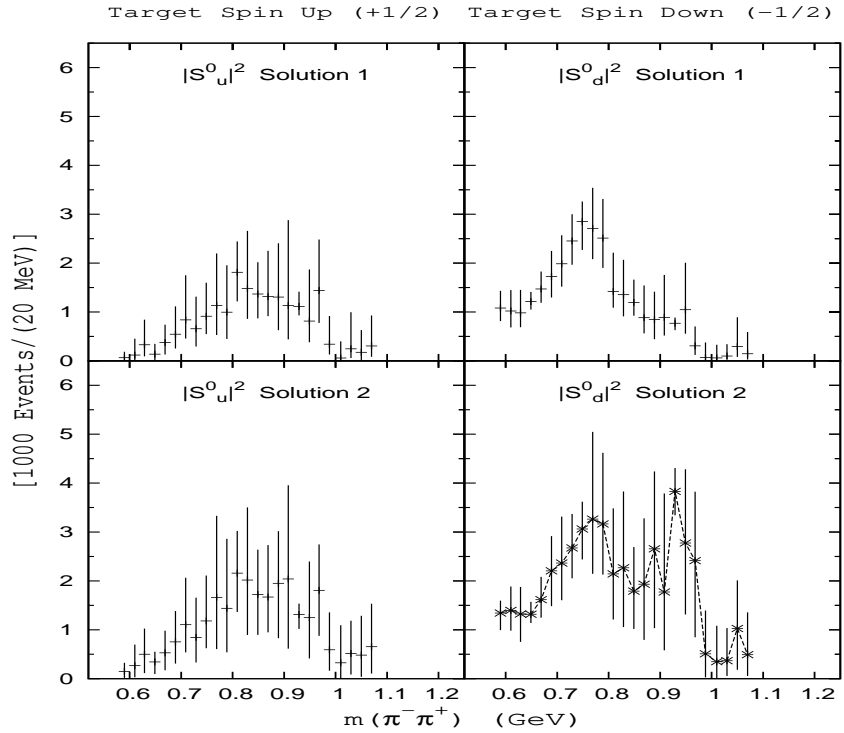}
\caption{Moduli of $S$-wave amplitudes $|S_\tau|^2$ from Analysis I in $\pi^- p \to \pi^- \pi^+ n$ at 17.2 GeV/c at low $t$~\cite{svec12b}.}
\label{Figure 17}
\end{figure}

\begin{figure} [htp]
\includegraphics[width=12cm,height=10.5cm]{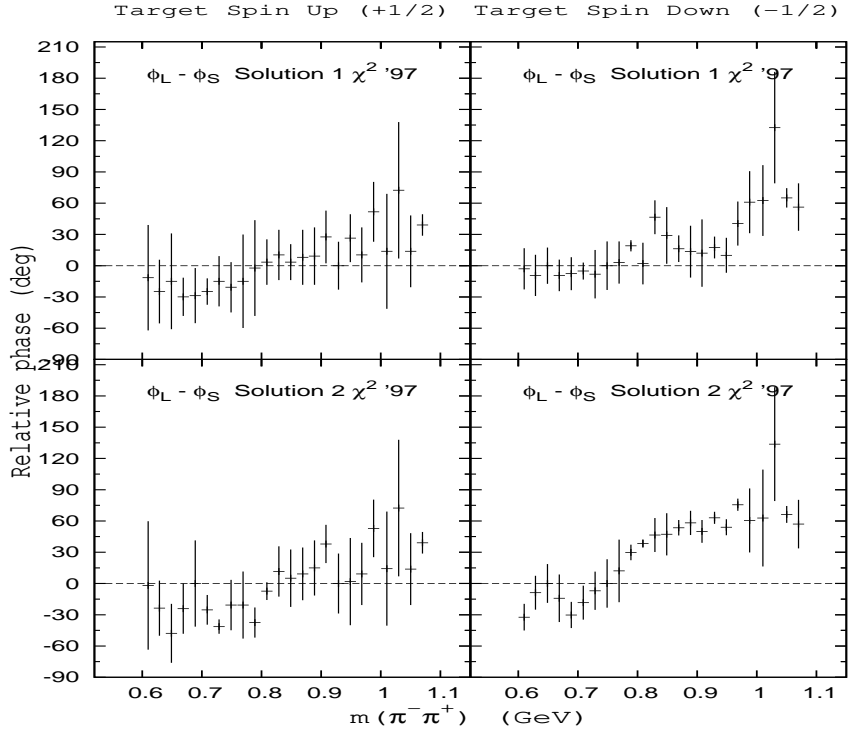}
\caption{Relative phases $\Phi(L_\tau)-\Phi(S_\tau)$ in $\pi^- p \to \pi^- \pi^+ n$ at 17.2 GeV/c at low $t$ from Ref.~\cite{kaminski97}.}
\label{Figure 18}
\end{figure}

\begin{figure}[hp]
\includegraphics[width=12cm,height=10.5cm]{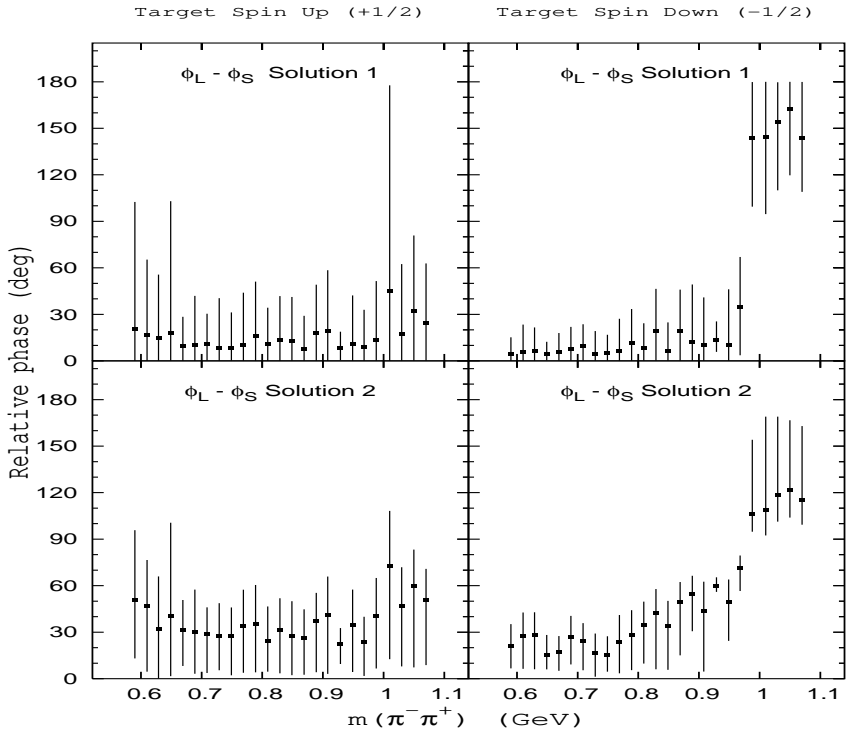}
\caption{Relative phases $\Phi_{L_\tau}-\Phi_{S_\tau}$ from Analysis I in $\pi^- p \to \pi^- \pi^+ n$ at 17.2 GeV/c at low $t$~\cite{svec12b}.}
\label{Figure 19}
\end{figure}

\begin{figure} [htp]
\includegraphics[width=12cm,height=10.5cm]{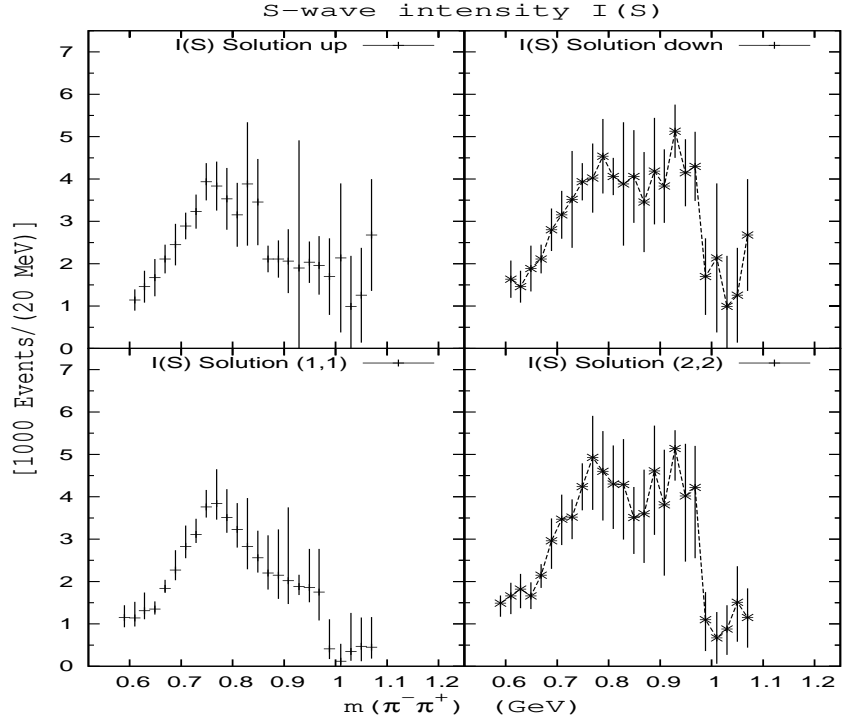}
\caption{Intensities $I(S)$ in $\pi^- p \to \pi^- \pi^+ n$ at 17.2 GeV/c from Ref.~\cite{kaminski97} (top) and Analysis I ~\cite{svec12b} (bottom).}
\label{Figure 20}
\end{figure}

\begin{figure} [htp]
\includegraphics[width=12cm,height=10.5cm]{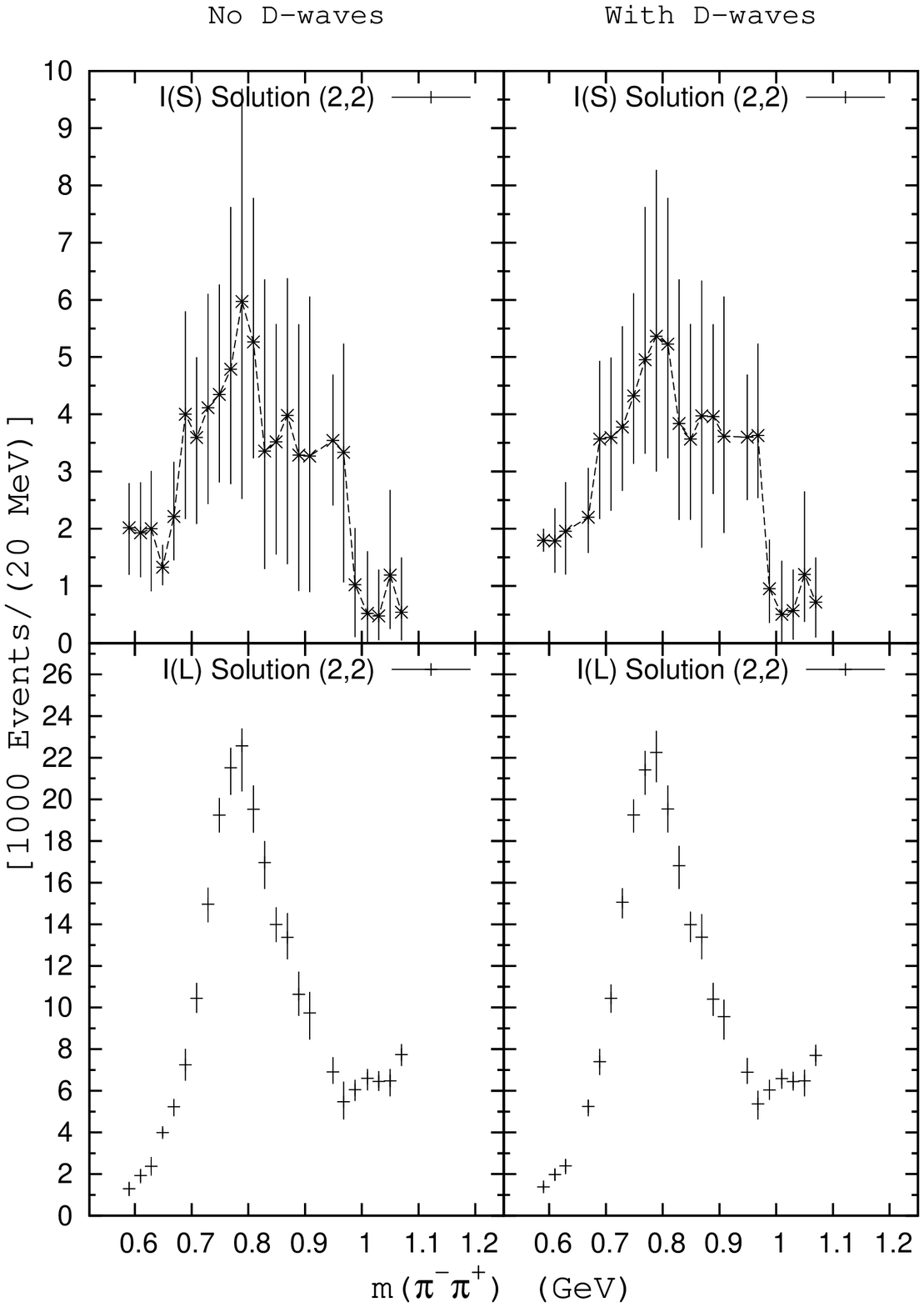}
\caption{Intensities in $\pi^- p \to \pi^- \pi^+ n$ at 17.2 GeV/c at low $t$ from analysis using spin mixing mechanism~\cite{svec14a}.}
\label{Figure 21}
\end{figure}

Figures 18 and 19 compare the relative phases 
$\Phi(L_\tau S^*_\tau)=\Phi(L_\tau)-\Phi(S_\tau)$ from the $\chi^2$ fit analysis~\cite{kaminski97} and Monte Carlo analysis~\cite{svec12b}, respectively. The $\chi^2$ fits suggest a zero structure of 
these relative phases near 700-800 MeV which allows for the change of sign of $\Phi(L_\tau)-\Phi(S_\tau)$ but does not require it. Such structure is in tension with the Monte Carlo Analysis I which allows for no change of sign of the phases. In fact, a detailed histogram analysis in steps of $1^\circ$ showed that below $5^\circ$ all degree bins for the Solution 2 of the phases $\Phi(L_u S^*_u)$ and $\Phi(L_d S^*_d)$ are empty for all mass bins except for just several events in a few mass bins below 800 MeV. The results of the $\chi^2$ fits are in tension also with our new analysis~\cite{svec14a} using the spin mixing mechanism~\cite{svec13b} which shows even a clearer non-zero structure of $\Phi(L_\tau S^*_\tau)$ at all dipion masses. 

The early analyses of the CERN data at low $t$~\cite{becker79b,chabaud83} were later updated in a new analysis by the Cracow group~\cite{kaminski97}. The results for the two solutions "up" and "down" for the $S$-wave intensity in the Cracow analysis are compared in Figure 20 with the corresponding two solutions (1,1) and (2,2) from the Monte Carlo analysis~\cite{svec12b}. The comparison shows that the Solution "up" is nearly identical to the Solution (1,1), and that the Solution "down" is nearly identical to the Solution (2,2). While the resonant structure at $\rho^0(770$ mass clearly peaks at 770 MeV in the Solution (2,2) it peaks at a lower value of the intensity at 790 MeV in the Solution "down". Despite this difference at 770 MeV both Solutions "down" and (2,2) communicate the convincing evidence for a  $\rho^0(770)$ structure in the Solution 2 of the transversity amplitudes $|S_d|^2$ seen in Figures 16 and 17.

While the Solutions "up"/(1,1) show a pronounced $\rho^0(770)$ structure this structure is less pronounced in the Solutions "down"/(2,2) due to a broad structure near 930 MeV originaing in the same structure in the Solution 2 for the amplitude $|S_d|^2$ in Figures 16 and 17. This structure is absent in Solution 2 for the amplitude $|S_d|^2$ in our new analysis~\cite{svec14a} using spin mixing mechanism in which the observed amplitudes are a certain mixture of $S$-matrix amplitudes with different spins~\cite{svec13b}. Spin mixing mechanism allows to extract information on the $D$-wave amplitudes and the amplitude analysis can be performed without and with the determination of the $D$-waves. Results for the Solution (2,2) of the $S$-wave and $P$-wave intensities $I(S)$ and $I(L)$ without and with $D$-wave determination are presented in Figure 21. The evidence for $\rho^0(770)$ in the Solution (2,2) of  $I(S)$ becomes clear. The sudden drop of $I(S)$ and the sudden rise of $I(L)$ at and above $f_0(980)$ mass is due to a strong mixing at these masses of $S$-wave and $P$-wave $S$-matrix amplitudes. Spin mixing mechanism excludes the Solution (1,1) of $I(S)$ and $I(L)$.

\subsection{A note on $\pi\pi$ phase-shift analyses of the $\pi^-\pi^+$ and $\pi^0\pi^0$ production data}

High statistics CERN-Munich data on $\pi^- p \to \pi^- \pi^+ n$ at 17.2 GeV/c on unpolarized target~\cite{grayer74} were analysed using several methods to determine $\pi\pi$ phase- 
shifts~\cite{hyams73,estabrooks73,estabrooks74,bugg96,martin76,petersen77}. First $\pi\pi$ phase-shift analysis using CERN-Cracow-Munich data on $\pi^- p \to \pi^- \pi^+ n$ at 17.2 GeV/c on polarized target was reported 1997 in Ref.~\cite{kaminski97} (henceforth referred to as KLR 97) and revisited 2002 in Ref.~\cite{kaminski02} (henceforth referred to as KLR 02). These analyses used model dependent methods to determine single flip helicity amplitudes from the measured transversity amplitudes. The helicity amplitudes were then related to $\pi \pi$ scattering amplitudes using pion exchange dominance approximation. 

In Figures 16 and 17 we have compared the two Solutions for $S$-wave transversity amplitudes $|S_u|^2$ and $|S_d|^2$ from the Cracow 1997 amplitude analysis $\chi^2$ $97$ with our amplitude Analysis I. The two analyses are nearly identical but make different assumptions about the helicity amplitudes. 
The $S$-wave helicity amplitude $|S_1|^2$ is given by~\cite{svec13a}
\begin{equation}
|S_1|^2=\frac{1}{2}(|S_u|-|S_d|e^{i\omega})^2
\end{equation}
where $\omega=\Phi(S_d)-\Phi(S_u)$ is the relative phase between $S$-wave transversity amplitudes of opposite transversity. The measurements on polarized target do not measure $\omega$ but at low $t$ it can be inferred from the self-consistecy conditions on the measured bilinear terms~\cite{svec12b}. Its exact value is $\omega=\pm180^\circ$. In our amplitude analysis the helicity amplitudes $|S_1|^2$ and $|L_1|^2$ are model independent and were determined in terms of the measured transversity amplitudes in Ref.~\cite{svec12b}. There are  two Solutions for our helicity amplitudes labeled (1,1) and (2,2).

The phase-shift analyses KLR $97$ and KLR $02$ both assume an Ansatz 
\begin{equation}
\omega = \Phi(S_dL_d^*)-\Phi(S_uL_u^*)+\Delta
\end{equation}
where $\Phi(S_\tau L_\tau^*)$ are the measured relative phases and $\Delta$ is a correction parameter. In the analysis KLR 97~\cite{kaminski97} $\Delta$ is a constant equal to $50.73^\circ$ below $K\bar{K}$ threshold. In the analysis KLR 02~\cite{kaminski02} $\Delta$ is a variable parameter. It was determined at each mass bin from the BNL data on $\pi^- p \to \pi^0 \pi^0 n$ at 18.3 GeV/c~\cite{gunter01} using a relation between the intensities $I$ and $I_0$ in $\pi^-\pi^+$ and $\pi^0 \pi^0$ production and the transversity amplitudes for both Solutions "up" and "down". 

In our recent work~\cite{svec15a} we perform two different $\pi\pi$ phase-shift analyses. The first analysis assumes elastic $\pi \pi$ scattering below the $K\bar{K}$ threshold. There are two analytical solutions for the phase shift $\delta^0_S$ for each input helicity amplitudes labeled (1,1)1, (1,1)2 and (2,2)1,(2,2)2. The physical Solution (2,2)1 is remarkably consistent with the 1997 Cracow Solution "down-flat" in KLR $97$.

The second analysis is a joint $\pi \pi$ phase shift analysis of the CERN $\pi^-\pi^+$~\cite{rybicki96} and E852 $\pi^0\pi^0$~\cite{gunter01,pi0pi0pwa} data. We obtain unique analytical solutions for the phase-shift $\delta^0_S$ and inelasticity $\eta^0_S$ for both input helicity amplitudes labeled (1,1) joint and (2,2) joint. Our results for $\delta^0_S$ in Solutions (1,1) joint and (2,2) joint are in excellent agreement with the 2002 Cracow Solutions "up-flat" and "down-flat" KLR $02$, respectively. Both our joint solutions have inelasticities $\eta^0_S <1$ in contrast to many unphysical values in KLR $02$.

The remarkable agreement of our elastic and joint Solutions with the "down-flat" Solutions KLR $97$ and KLR $02$, respectivey indicates that the differences in the relative phase $\omega$ play a minor role in the determination of the phase-shift $\delta^0_S$ but can affect the determination of the inelasticity. Our central observation is that all these Solutions for phase-shift $\delta^0_S$ are fully consistent with the evidence for $\rho^0(770)-f_0(980)$ spin mixing in the measured transversity amplitudes from which all these phase-shifts ultimately arise.

\section{Summary and Outlook.}

We have studied the response of the analytical solutions of the $S$- and $P$-wave subsystem in $\pi^- p \to \pi^- \pi^+ n$ on polarized target to the presence of $D$-wave amplitudes. We have found that the $\rho^0(770)-f_0(980)$ mixing is consistent with the presence of $D$-wave amplitudes with helicities $\lambda \leq 1$ (Response analysis A) as well as with helicities $\lambda \leq 2$ (Response analysis B) provided the $D$-wave amplitudes are not too large below 750 MeV, as expected from the CERN measurements on unpolarized and polarized targets. Above 750 MeV the spin mixing effect is consistent with larger $D$-waves amplitudes. This result is in agreement with the amplitude analysis of $S$, $P$ and $D$ wave subsystem at high momentum transfer $t$ below 960 MeV. The observed $\rho^0(770)-f_0(980)$ mixing is thus a real effect not generated by the small $D$-wave contamination of the input data. We have also shown that the spin mixing $S$-wave amplitudes in $\pi^- p \to \pi^- \pi^+ n$ are consistent with isospin relations between the $S$-wave amplitudes in $\pi^-\pi^+$, $\pi^0 \pi^0$ and $\pi^+ \pi^+$ channels and thus with the data on $\pi^- p \to \pi ^0 \pi^0 n$ and $\pi^+ p \to \pi ^+ \pi^+ n$. 

Next we have presented a complete survey of evidence for $\rho^0(770)-f_0(980)$ mixing. We conclude that all amplitude analyses of all five measurements on polarized targets show a clear evidence for the $\rho^0(770)-f_0(980)$ mixing in the $S$-wave moduli and intensities. Apart from the tension in the zero structure of relative phases $\Phi_{L}-\Phi_{S}$ near 700-800 MeV in our Analysis I~\cite{svec12b} and the analysis~\cite{kaminski97} the relative phases in these analyses have similar magnitudes. The tension in the zero structure of $\Phi_{L}-\Phi_{S}$ may reflect the fact that while in our analysis the cosine conditions are imposed on solutions for the amplitudes for every Monte Carlo sampling of the data error volume they are not imposed on the amplitudes by the $\chi^2$ minimization program. This difference appears significant only near 700-800 MeV.

Finally, we have commented on our elastic and joint $\pi\pi$ phase-shits analyses of the $\pi^-\pi^+$ and $\pi^0\pi^0$ data~\cite{svec15a} and their agreement with the Cracow Solutions~\cite{kaminski97,kaminski02}. All solutions for $\delta^0_S$ are consistent with the evidence for $\rho^0(770)-f_0(980)$ mixing surveyed in the Section VI.

The consistency of $\rho^0(770)-f_0(980)$ mixing with the presence of $D$-wave amplitudes, with the amplitude analysis of $\pi^-p \to \pi^0\pi^0n$ BNL data and the mutual consistency of all amplitude analyses on polarized target and all $\pi\pi$ phase-shift analyses are results that significantly strengthen the experimental evidence for $\rho^0(770)-f_0(980)$ spin mixing~\cite{svec12b}. This evidence is further supported by our new amplitude analysis~\cite{svec14a} using spin mixing mechanism developed in Ref.~\cite{svec13a,svec13b}. The origin of $\rho^0(770)-f_0(980)$ spin mixing is in a new non-standard interaction of particle scattering processes with a quantum environment~\cite{svec13a,svec13b} which we propose to identify with dark matter~\cite{svec14a}.

\end{document}